\documentclass[lettersize,journal]{IEEEtran}
\usepackage{fancyhdr}

\usepackage{amsmath} 
\usepackage{amssymb} 
\usepackage{indentfirst}
\usepackage{graphicx}
\IEEEoverridecommandlockouts
\usepackage{cite}
\usepackage{amsmath,amssymb,amsfonts}
\usepackage{algorithmicx}
\usepackage{algorithm}
\usepackage{algpseudocode}  
\usepackage{amsmath}  
\usepackage{graphicx}
\usepackage{textcomp}
\usepackage{xcolor}
\usepackage{float}
\usepackage{graphicx}
\usepackage[caption=false,font=footnotesize]{subfig}
\usepackage{verbatim}
\def\BibTeX{{\rm B\kern-.05em{\sc i\kern-.025em b}\kern-.08em
    T\kern-.1667em\lower.7ex\hbox{E}\kern-.125emX}}

\usepackage{xcolor}
\usepackage{xpatch}
\makeatletter
\def\changeBibColor#1{%
  \in@{#1}{}
  \ifin@\color{blue}\else\normalcolor\fi
}

\xpatchcmd\@bibitem
  {\item}
  {\changeBibColor{#1}\item}
  {}{\fail}
 
\xpatchcmd\@lbibitem
  {\item}
  {\changeBibColor{#2}\item}
  {}{\fail}
\makeatother
\begin{document}
\title{ Polarization-aware Reconfigurable Antenna Aided Wireless Communications}

\author{ 
Chao Zhang, \emph{Graduate Student Member, IEEE}, Hu Zhou, \emph{Member, IEEE}, Ruizhe Long, \emph{Member, IEEE},\\  Ying-Chang Liang, \emph{Fellow, IEEE}, and Boon‑Hee Soong, \emph{Life Senior Member, IEEE}
\\
\thanks{C.~Zhang, H.~Zhou and R.~Long are with the National Key Laboratory of Wireless Communications, University of Electronic Science and Technology of China, Chengdu 611731, China (e-mail: {zhang\_chao@std.uestc.edu.cn, huzhou@std.uestc.edu.cn, ruizhelong@gmail.com}).}
\thanks{Y.-C. Liang is with the Institute for Fundamental and Frontier Sciences, University of Electronic Science and Technology of China, Chengdu 611731, China (e-mail: {liangyc@ieee.org}). }
\thanks{B. H. Soong is with the Department of Electrical and Electronic Engineering, Nanyang Technological University, Singapore 639798 (e-mail: {ebhsoong@ntu.edu.sg}).}
}
\maketitle
\thispagestyle{fancy}
\lhead{This work has been submitted to the IEEE for possible publication. Copyright may be transferred without notice, after which this version may no longer be accessible.}

\begin{abstract}
Reconfigurable antennas (RAs) have emerged as a promising technology for future wireless networks, offering additional flexibility for wireless communications. Among existing designs, rotatable antennas are particularly effective in improving directional gain via boresight alignment only. However, conventional rotatable RAs often overlook a critical physical coupling: the mechanical rotation inevitably alters the radiated polarization orientation, potentially leading to polarization mismatch. To address this challenge, in this paper, we investigate a novel RA architecture that simultaneously supports three-dimensional rotation and polarization state reconfiguration, ensuring alignment in both spatial and polarization domains.
To quantify the performance gains unlocked by addressing this mismatch, we analyze a simplified single-user LoS scenario to compare the optimized rotatable design against a fixed scheme.
This analysis explicitly attributes the performance improvement to three aspects: the directional and projection gain arising from boresight steering, the polarization direction alignment gain enabled by roll adjustment, and the polarization state matching gain provided by polarization reconfiguration.
Furthermore, for general multipath multi-user systems, we formulate a joint power minimization problem by optimizing the digital beamforming alongside the rotation and polarization designs, subject to rate requirements and hardware constraints. To solve the resulting non-convex problem, we develop an alternating optimization framework to obtain a solution efficiently, where the digital beamforming is solved via semidefinite relaxation and difference-of-convex techniques, while the rotation and polarization designs are updated using Riemannian conjugate gradient on their respective manifolds. Simulation results demonstrate that the proposed RA outperforms both rotation-only and boresight-only benchmarks, achieving lower transmit power under the same rate constraints by joint spatial-polarization design. 
\end{abstract}


\section{Introduction}
Emerging $6$G applications, such as immersive XR, digital twins, and connected and autonomous vehicles, impose stringent requirements on peak data rates, reliability, and energy efficiency~\cite{6G1,6G2,6G3}. To meet these demands, multi-antenna techniques have evolved from multi-input multi-output (MIMO) to massive MIMO and, more recently, extremely large-scale MIMO~\cite{MIMO1,bjornson2025enabling,zhang_FTTD}. While these advancements have significantly expanded system capacity, further scaling up array sizes faces practical bottlenecks, such as limited physical aperture and strong spatial correlation, which may reduce the effective spatial degrees of freedom (DoFs). These limitations motivate the pursuit of new controllable DoFs at the physical layer, beyond merely scaling up array size.

To address these challenges, reconfigurable antennas (RAs) have attracted increasing attention as a hardware-efficient paradigm that introduces additional controllable DoFs in the antenna domain.
Depending on the implementation, RAs can reconfigure antenna geometry (position and orientation), polarization state, and even frequency response~\cite{MA1,MA2,6DMA,PandF}.

Among the various RA capabilities, geometry reconfiguration has received considerable attention for its ability to adaptively reshape propagation channels. Initial studies focused on movable antenna (MA) and fluid antenna system (FAS)~\cite{MABF1,MABF2,MABF3,AnJC}, where the physical locations of antenna elements are optimized within a feasible region to exploit location-dependent channel variations and improve the effective channel quality. By changing path lengths and phases, position adjustment reshapes the effective channel and can alleviate severe fades while enhancing spatial separability across users.
Specifically, the energy-efficiency maximization problem for near-field FAS was studied in~\cite{MABF1} by jointly optimizing BS transmit beamforming and antenna positions via an alternating optimization (AO) algorithm, showing notable performance gains and validating the effectiveness of FAS. In \cite{MABF2}, an MA-aided multi-user hybrid beamforming scheme was proposed, where an AO algorithm was developed to jointly design beamforming and subarray positions to boost the sum rate over fixed-position arrays. Moreover, the MA beamforming was proposed in~\cite{MABF3} by jointly optimizing antenna positions and weights, enabling full array gain toward the desired direction and null steering toward all undesired directions under certain conditions, with closed-form solutions and validated performance gains over fixed-position arrays.

While these results confirm that physically reconfiguring antenna locations effectively reshapes the channel, position adjustment is only one aspect of geometric reconfigurability.
Beyond position adjustment, another practical option is antenna orientation, which effectively extends the service coverage from conventional terrestrial areas to the full three-dimensional (3D) space, empowering the burgeoning low-altitude economy by ensuring robust link quality for aerial users (e.g., UAVs) with highly dynamic elevation angles~\cite{10972091}. 
Regarding rotatable antennas, a rotatable model was proposed in~\cite{RAA1} to enhance uplink communications by jointly optimizing the BS receive beamforming and the 3D boresight directions of RAs to maximize the minimum signal-to-interference-plus-noise ratio (SINR). To mitigate interference, a rotatable antenna array was proposed in~\cite{RAA2} to enhance null steering.
In this scheme, the entire array was rotated to modify the effective position and orientation of antennas, and the rotation angles were optimized to maximize the gain in the desired direction while nulling interference.
Additionally, a rotatable antenna-assisted cognitive radio system was proposed in~\cite{RAA4}, achieving notable spectrum-sharing gains by jointly optimizing secondary transmit beamforming and RA boresight directions.

However, these rotatable designs fundamentally overlook a critical physical coupling: in 3D space, rotating an antenna steers its boresight and simultaneously rotates its polarization orientation, which may lead to polarization mismatch. 
Even with perfect spatial alignment, signal power cannot be effectively harvested if the polarization states are orthogonal~\cite{balanis2016antenna}. In practice, the arriving polarization is highly unpredictable due to random device orientations and channel depolarization, where mechanisms like reflection and scattering alter the incident polarization state. Consequently, conventional systems relying on fixed polarization often suffer from severe polarization mismatch loss. Although dual-polarized or even tri-polarized antennas offer improved robustness~\cite{nature}, they typically necessitate additional radio frequency (RF) chains, leading to higher hardware costs and design complexity. This challenge has motivated extensive research on polarization reconfigurable antennas, which can adaptively tune their polarization state to match the channel with modest implementation overhead. Specifically, polarization reconfigurable antennas utilizing discrete selection among a small set of predefined states were investigated in \cite{PRA1,PRAA}, offering a practical approach to switch polarization with limited hardware overhead. Furthermore, polarization-agile designs were considered in \cite{PRA3} to enable continuous adjustment of the polarization orientation, providing finer adaptation to channel variations. More recently, a phase shifter based polarization reconfigurable design was proposed in \cite{PRA4}, where the relative phase between orthogonal polarization components is controlled to synthesize different polarization states, realizing flexible polarization adaptation with low implementation cost.

Overall, geometric reconfiguration and polarization reconfiguration have generally been treated as two separate design variables in most existing works, failing to address their inherent coupling mentioned above. The physical basis of this coupling was investigated in~\cite{Joint1}, which verified that the spatial geometry of the transceiver setup dictates the polarization characteristics. However, this study was limited to channel modeling and treated antenna orientation as a static deployment configuration rather than active DoFs for system optimization. Building on this geometric coupling, joint rotation and polarization optimization was recently explored in \cite{Joint2} for a 6DMA-enabled ISAC system. Nevertheless, this work focused on localization applications within LoS settings. In this 6DMA context, rotation is utilized primarily to exploit spatial DoFs, rather than exploiting the high directional gain provided by antenna steering. Conversely, rotatable directional antennas are characterized by high directivity. In these systems, existing works primarily focus on boresight alignment, demonstrating that higher antenna gain generally leads to superior performance. However, prioritizing boresight steering solely to maximize directional gain overlooks the inherent coupling between rotation and polarization. Consequently, without effective polarization matching, the advantage of high directional gain is compromised.


Motivated by the above studies and the remaining gaps, we propose a RA-aided system model in which each antenna element can rotate in 3D space and electronically reconfigure its polarization state, as illustrated in Fig.~\ref{figure} and Fig.~\ref{fig:overall}. By jointly exploiting 3D rotation for directional gain enhancement and geometric polarization direction alignment, augmented by polarization reconfiguration to flexibly adapt to channel depolarization, the proposed architecture provides additional flexibilities beyond conventional digital beamforming. Specifically, we first consider a simplified single-user LoS scenario to explicitly attribute the performance improvements to directional and polarization matching gains. Subsequently, we extend the framework to multi-user multipath systems. Recognizing that achieving directional and polarization matching gains is no longer straightforward due to severe channel depolarization and inter-user interference, we formulate a joint optimization problem and develop an efficient solution.
The main contributions of this paper are summarized as follows:

\begin{itemize}
  \item We propose a RA architecture that enables {3D rotation} together with {polarization reconfiguration} at each antenna element to provide additional DoFs for transmission enhancement beyond conventional digital beamforming.
  \item To reveal the benefits of the proposed RA, we provide theoretical performance analysis for a simplified single-user LoS setting. We explicitly decompose the performance improvement into directional gain and polarization matching gain. Furthermore, we reveal that the polarization matching gain originates from projection gain, polarization direction alignment, and polarization state matching.
  \item For general multipath channels and multi-user transmission, we formulate a joint optimization problem that minimizes the total transmit power subject to users' rate requirements and practical hardware constraints on rotation and polarization reconfiguration. To handle the resulting non-convexity, we develop an AO algorithm, where the digital beamforming subproblem is solved via semidefinite programming (SDP) combined with difference-of-convex (DC) techniques, and the rotation and polarization variables are updated efficiently using Riemannian conjugate gradient (RCG) on the corresponding manifolds.
  \item Simulation results demonstrate that the proposed RA with jointly adjustable rotation and polarization consistently outperforms baseline schemes with rotation-only and boresight-only RAs, achieving lower transmit power under the same rate constraints and improved overall system performance.
\end{itemize}

The rest of the paper is organized as follows: Section \uppercase\expandafter{\romannumeral2} introduces the system model of the proposed RA-enabled MU-MISO system. Section \uppercase\expandafter{\romannumeral3} provides a theoretical performance analysis for a simplified single-user LoS scenario. Section \uppercase\expandafter{\romannumeral4} formulates the joint digital beamforming, rotation, and polarization optimization problem for power minimization. Section \uppercase\expandafter{\romannumeral5} develops an AO algorithm based on SDP and RCG methods. Simulation results are presented in Section \uppercase\expandafter{\romannumeral6}, and Section \uppercase\expandafter{\romannumeral7} concludes the paper.

Notations: Lowercase and uppercase bold letters represent vectors and matrices, respectively. The conjugate, transpose and conjugate transpose of matrix $\bf A $ are denoted by ${\bf A}^* $, ${\bf A}^T $ and ${\bf A}^H $, respectively. $\mathbb{C}^{M \times N}$ and $\mathbb{R}^{M \times N}$ denote the spaces of $M \times N$ complex and real matrices, respectively. $\mathbf{A} \succeq \mathbf{0}$ indicates that $\mathbf{A}$ is a positive semi-definite (PSD) matrix. $\mathrm{Tr}(\mathbf{A})$ and $\|\mathbf{A}\|_2$ denote the trace and spectral norm of matrix $\mathbf{A}$, respectively. For a vector $\mathbf{a}$, $\|\mathbf{a}\|$ denotes its Euclidean norm, and $|[\mathbf{a}]_i|$ denotes the modulus of its $i$-th element. The symbol $\odot$ represents the Hadamard product, and $\mathfrak{R}\{\cdot\}$ denotes the real part of a complex value. Finally, $\mathcal{CN}(\mu, \sigma^2)$ represents the circularly symmetric complex Gaussian (CSCG) distribution with mean $\mu$ and variance $\sigma^2$.

\section{System Model}
\begin{figure}[t!]
  \centering
  \includegraphics[width=0.48\textwidth]{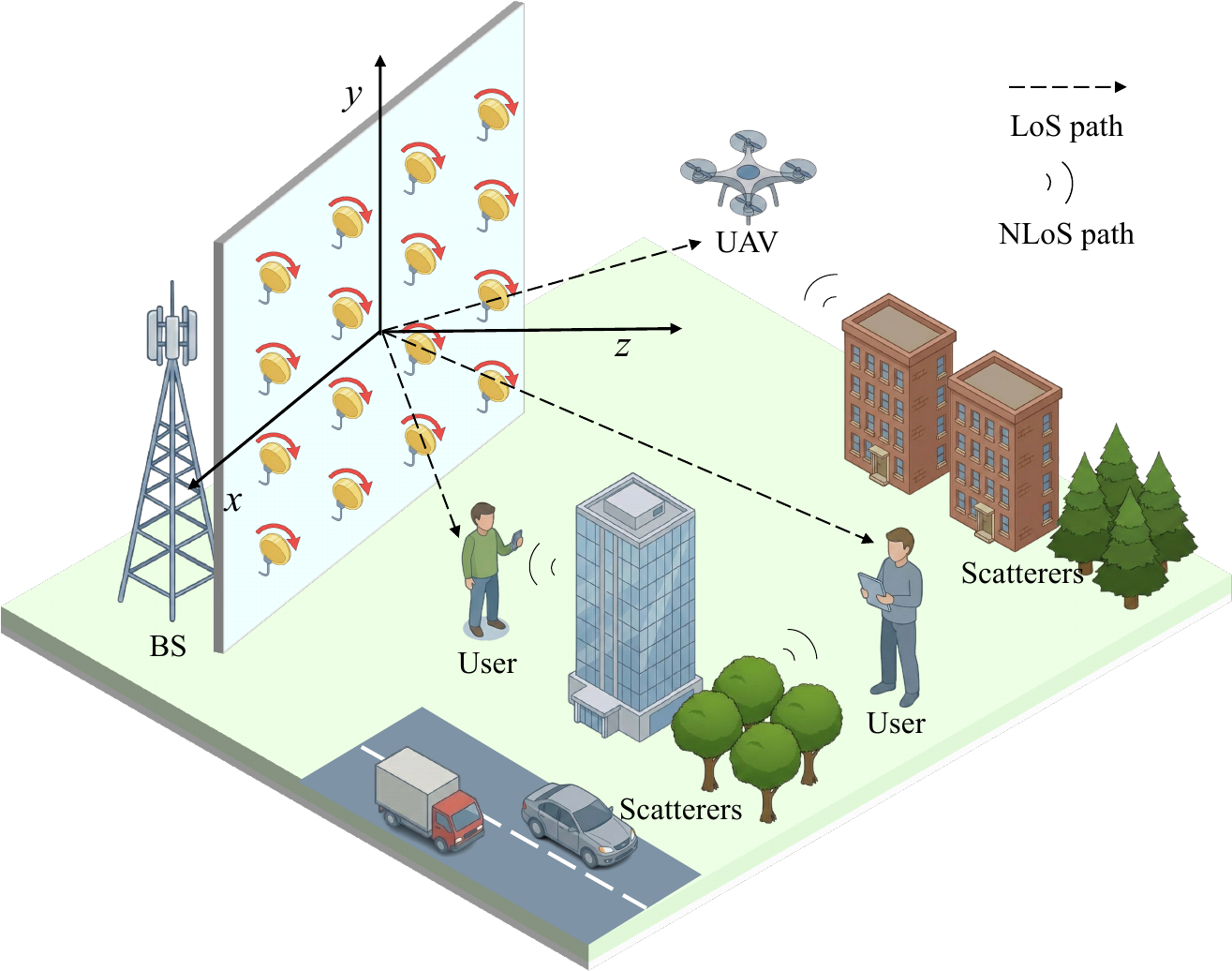}
  \caption{The RA-aided communication system}
  \label{figure}
\end{figure}

As illustrated in Fig. \ref{figure}, we consider a downlink multi-user MISO system where a base station (BS) serves $K$ users. The BS is equipped with $M$ rotatable antennas with reconfigurable polarization, arranged as a uniform planar array (UPA) in the $x$-$y$ plane. To enable polarization reconfigurability, each antenna element at the BS comprises two co-located orthogonally oriented linearly polarized dipoles: one vertically polarized (V-element) and one horizontally polarized (H-element). Similarly, at the user side, each user is equipped with a dual-polarized antenna composed of two orthogonal elements (H/V ports) with omnidirectional radiation patterns. 
Specifically, the BS employs a UPA with $M=M_xM_y$ antennas. For the $(n_x,n_y)$-th antenna, where $n_x\in\{0,1,...,M_x-1\}$, $n_y\in\{0,1,...,M_y-1\}$, its location in the global Cartesian coordinate system (CCS) is given by 
\begin{align}
  {\bf p}_{n_x,n_y}=\left[\begin{array}{c}
\left(n_x-\frac{M_x-1}{2}\right) \Delta  \\\left(n_y-\frac{M_y-1}{2}\right) \Delta\\
0
\end{array}\right], \forall n_x,n_y,\tag{1} \label{1}
\end{align}
where $\Delta$ denotes the spacing between adjacent antennas. For notational convenience, we map $(n_x,n_y)$ to a single index $m\in\{1,\ldots,M\}$ and denote the corresponding position by ${\bf p}_m$. \looseness-1
\subsection{RA Architecture and Modeling}
We consider a RA architecture capable of adjusting both its 3D rotation and polarization states.
\subsubsection{3D Rotation Model}
Let the global orthonormal basis be defined as ${\bf e}_x=[1,0,0]^T$, ${\bf e}_y=[0,1,0]^T$, and ${\bf e}_z=[0,0,1]^T$. For the $m$-th transmit antenna, we define the unrotated vertical-polarization direction as ${\bf e}_y$, the horizontal-polarization direction as ${\bf e}_x$, and the antenna boresight direction as ${\bf e}_z$. 
Let $\mathbf{R}_{m}=[{\bf r}_{m,1},{\bf r}_{m,2},{\bf r}_{m,3}]$ denote the rotation matrix for the $m$-th antenna. Mathematically, $\mathbf{R}_{m}$ belongs to the special orthogonal group $\mathrm{SO}(3)$, defined as
\begin{equation}
    \mathrm{SO}(3) \triangleq \{ \mathbf{R} \in \mathbb{R}^{3 \times 3} \mid \mathbf{R}^T \mathbf{R} = \mathbf{I}_3, \det(\mathbf{R}) = 1 \}.\tag{2}\label{2}
\end{equation}
The constraint $\mathbf{R}_{m} \in \mathrm{SO}(3)$ enforces a rigid-body rotation, preserving the orthogonality between the polarization ports and the boresight.
After rotation, the polarization directions become ${\bf e}_{H,m}=\mathbf{R}_{m}{\bf e}_x={\bf r}_{m,1}$, ${\bf e}_{V,m}=\mathbf{R}_{m}{\bf e}_y={\bf r}_{m,2}$, while the rotated boresight direction becomes ${\bf e}_{B,m}=\mathbf{R}_{m}{\bf e}_z={\bf r}_{m,3}$. 
In essence, the third column ${\bf r}_{m,3}$ determines the {boresight direction}, while the first two columns $\{{\bf r}_{m,1}, {\bf r}_{m,2}\}$ specify the H/V polarization directions in the 3D space.
\subsubsection{Polarization Reconfiguration Model}
\begin{figure}[t!]
\centering

\subfloat[Transmitter architecture\label{fig:subfig1}]{
    \includegraphics[width=0.45\textwidth]{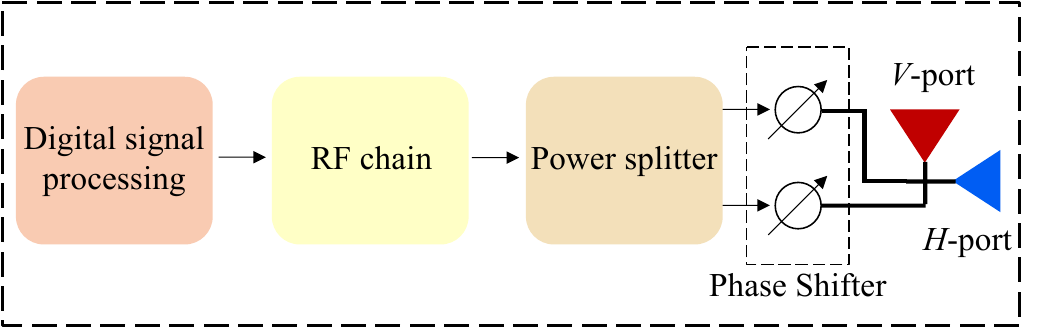}
}\\[0.6em]

\subfloat[Receiver architecture\label{fig:subfig2}]{
    \includegraphics[width=0.3\textwidth]{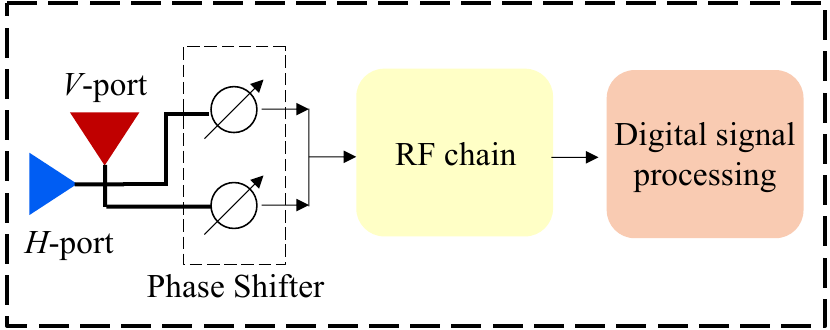}
}

\caption{Architecture of the polarization reconfigurable antenna}
\label{fig:overall}
\vspace{-1em}
\end{figure}

We next describe the polarization reconfiguration architecture, as illustrated in Fig.~\ref{fig:overall}. 
Each RA element is equipped with two {co-located} orthogonal polarization ports on the same physical antenna. 

\emph{Transmitter Side (BS)}: In contrast to conventional fixed-polarization configurations, the output of the $m$-th RF chain is connected to the two polarization ports via an adjustable power splitter, followed by two independent phase shifters. 
The polarization state vector of the $m$-th antenna, denoted by ${\bf{v}}_m \in \mathbb{C}^{2\times 1}$, is modeled as
\begin{align}
  {\bf v}_m=\left[\begin{array}{c}
\rho_{H,m}e^{-j\varphi_{H,m}}  \\ \rho_{V,m}e^{-j\varphi_{V,m}}
\end{array}\right], \forall m,\tag{3} \label{3}
\end{align}
where $\rho_{H,m}, \rho_{V,m}\ge 0$ represent the amplitude coefficients for the horizontal and vertical ports, respectively, satisfying the power constraint $\rho_{H,m}^2+\rho_{V,m}^2=1$. Moreover, the parameters $\varphi_{H,m},\varphi_{V,m}\in (0,2\pi]$ represent the phase shifts introduced by the phase shifters. 

\emph{Receiver Side (User)}: At the user side, to reduce hardware complexity, we consider a phase-only control architecture. Specifically, the signals from the two orthogonal receive ports are phase-shifted and then combined into a single RF chain. For user $k$, the receive polarization combining vector is defined as:
${\bf u}_k=\left[\begin{array}{c}
e^{-j\phi_{H,k}}  \\ e^{-j\phi_{V,k}}
\end{array}\right], \forall k$, where $\phi_{H,k},\phi_{V,k}\in (0,2\pi]$ represent the phase shifts introduced by the phase shifters. The proposed architecture allows both the transmitter and receiver to synthesize various polarization states, such as linear, elliptical, and circular polarizations, by adjusting the phase difference. This flexibility enables the system to effectively compensate for polarization mismatch.\looseness-1
\subsection{Channel Model}
We consider a geometry-based channel model consisting of one LoS path and $L$ NLoS scattering paths~\cite{he2022polarization,zhu20213gpp}. Let ${\bf p}_k \in \mathbb{R}^{3 \times 1}$ denote the location of user $k$, and let ${\bf s}_l \in \mathbb{R}^{3 \times 1}$ denote the location of the $l$-th scatterer. Thus, the downlink channel coefficient between the $m$-th transmit antenna and the $k$-th user is denoted by $h_{m,k}$. Specifically, the overall channel coefficient is obtained by summing the contributions of the LoS path and the $L$ NLoS paths, i.e.,
\begin{align}
  h_{m,k}=\sum_{l=0}^L h_{m,k,l},\tag{4} \label{4}
\end{align}
where $h_{m,k,0} $ denotes the LoS component, and $h_{m,k,l}$ denotes the $l$-th NLoS component for $l=1,\ldots,L$.

\subsubsection{LoS Path Modeling} For the LoS path, the distance between the $m$-th antenna and user $k$ is  $d_{m,k,0}=\|{\bf p}_k-{\bf p}_m\|$. The corresponding unit propagation direction vector is given by ${{\bf f}_{m,k,0}}=\frac{{\bf p}_k-{\bf p}_m}{\|{\bf p}_k-{\bf p}_m\|}$. We adopt the far-field assumption, where the radiated electric field lies in the transverse plane orthogonal to the propagation direction. We construct an orthonormal basis
${\bf Z}_{m,k,0}=[{\bf z}_{m,k,0},\bar{\bf z}_{m,k,0}]\in \mathbb{R}^{3 \times 2}$ for the transverse plane, satisfying ${\bf Z}_{m,k,0}^T{\bf Z}_{m,k,0}={\bf I}_2$ and ${\bf Z}_{m,k,0}^T{{\bf f}_{m,k,0}}={\bf 0}$. Let ${\bf E}=[{\bf e}_x,{\bf e}_y]\in \mathbb{R}^{3 \times 2}$ denote the reference basis of the H/V ports in the global CCS. Accordingly, by projecting the rotated antenna basis $\mathbf{R}_m \mathbf{E}$ onto the defined transverse plane, the transmit polarization projection matrix is given by ${\bf P}_{m,k,0}={\bf Z}_{m,k,0}^T{\bf R}_m{\bf E}\in \mathbb{R}^{2 \times 2}$. Similarly, the receive polarization projection matrix is defined as ${\bf Q}_{m,k,0}={\bf E}^T{\bf Z}_{m,k,0}\in \mathbb{R}^{2 \times 2}$, since the user employs non-rotatable antennas aligned with the global CCS. 
Thus, the small-scale component accounting for polarization matching is given by ${\bf u}^H_k{\bf Q}_{m,k,0}{\bf M}_{m,k,0}{\bf P}_{m,k,0}{\bf v}_m$, where ${\bf M}_{m,k,0}$ denotes the polarization coupling matrix. For the LoS component, we assume polarization preservation. Hence, the corresponding polarization coupling matrix is modeled as the identity matrix, i.e., ${\bf M}_{m,k,0}={\bf I}_2$.

For the large-scale component, we consider directional antennas at the BS. The effective gain depends on the misalignment angle between the antenna boresight and the propagation direction. Specifically, for the LoS path, the misalignment angle is
\begin{align}
  \varepsilon_{m,k,0}=\arccos\left({\bf e}_{B,m}^T{{\bf f}_{m,k,0}}\right),\tag{5} \label{5}
\end{align}
where $\mathbf{e}_{B,m} = \mathbf{R}_m \mathbf{e}_z$ is the current boresight direction. We adopt the widely used cosine-power pattern
\begin{align*}
  G(\varepsilon)=\left\{\begin{array}{ll}
G_0 \cos ^{2 p}(\varepsilon), & \varepsilon \leq \pi / 2, \\
0, & \text { otherwise},
\end{array} \right. \tag{6} \label{6}
\end{align*}
where $p\ge 0$ denotes the directivity factor and $G_0=2(2 p+1)$ denotes the maximum gain.The path loss factor is given by the Friis transmission equation:
\begin{align*}
  \beta_{m,k,0}\approx \frac{A}{4\pi d_{m,k,0}^2}G_0 \cos ^{2 p}(\varepsilon_{m,k,0}),\tag{7} \label{7}
\end{align*}
where $A$ denotes the physical size of each antenna. Since the user antenna is assumed omnidirectional, the receive-side gain is set to unity. Thus, the channel $h_{m,k,0}$ can be expressed as 
\begin{align*}
  h_{m,k,0}\!\!=\!\!\sqrt{\!\beta_{m,k,0}}{\bf u}^H_k\!{\bf Q}_{m,k,0}{\bf M}_{m,k,0}{\bf P}_{m,k,0}{\bf v}_m e^{-j\!\frac{2\pi}{\lambda}\!d_{m,k,0}}\!.\tag{8} \label{8}
\end{align*}

\subsubsection{NLoS Path Modeling} Similarly, the channel coefficient of the $l$-th NLoS path for all $l\ge 1$  is expressed as
\begin{align*}
  h_{m,k,l}\!=\sqrt{\beta_{m,k,l}}{\bf u}^H_k{\bf Q}_{k,r,l}{\bf M}_{m,k,l}{\bf P}_{m,t,l}{\bf v}_m e^{-j\frac{2\pi}{\lambda}d_{m,k,l}},\tag{9} \label{9}
\end{align*}
where $\beta_{m,k,l}$ and $d_{m,k,l}$ denote the path loss and the total propagation distance, respectively. Unlike the LoS path, the NLoS component involves scattering. Let $\mathbf{f}_{m,l}^{\text{tx}} = \frac{\mathbf{s}_l - \mathbf{p}_m}{\|\mathbf{s}_l - \mathbf{p}_m\|}$ denote the departure direction from the $m$-th BS antenna to scatterer $l$, and $\mathbf{f}_{k,l}^{\text{rx}} = \frac{\mathbf{p}_k - \mathbf{s}_l}{\|\mathbf{p}_k - \mathbf{s}_l\|}$ denote the arrival direction from scatterer $l$ to user $k$. Accordingly, we construct the transverse plane bases $\mathbf{Z}_{m,l}^{\text{tx}}$ and $\mathbf{Z}_{k,l}^{\text{rx}}$ for these directions. The transmit and receive projection matrices are then given by ${\bf P}_{m,l}=(\mathbf{Z}_{m,l}^{\text{tx}})^T{\bf R}_m{\bf E}$ and ${\bf Q}_{k,l}={\bf E}^T\mathbf{Z}_{k,l}^{\text{rx}}$, respectively. The polarization coupling matrix $\mathbf{M}_{m,k,l}$ is determined by the electromagnetic properties of the scatterer, such as its material and geometry. It is modeled as
\begin{align*}
  {\bf M}_{m,k,l}=\left[\begin{array}{ll} \sqrt{\chi}e^{j\theta^{HH}_{m,k,l}} & \sqrt{1-\chi}e^{j\theta^{HV}_{m,k,l}} \\ \sqrt{1-\chi}e^{j\theta^{VH}_{m,k,l}} & \sqrt{\chi}e^{j\theta^{VV}_{m,k,l}} \end{array}\right],\tag{10} \label{10}
\end{align*}
where $\chi\in [0,1]$ represents the co-polarization power fraction and $\theta^{ab}\in(0,2\pi]$ for $a,b \in\{H,V\}$ represent the independent random phase shifts from transmit polarization $b$ to receive polarization $a$ introduced by the scatterer \cite{zhu20213gpp}. The diagonal terms correspond to co-polar responses, while the off-diagonal terms model cross-polar coupling introduced by scattering. Here, we define the cross-polarization discrimination (XPD) as the ratio between co-polar and cross-polar power, i.e., $\text{XPD}=\frac{\chi}{1-\chi}$. Hence, a larger XPD corresponds to smaller cross-polar leakage.

Finally, we stack $\{h_{k,m}\}_{m=1}^{M}$ into the downlink channel vector for user $k$, denoted as 
${\bf h}_k=[h_{1,k},h_{2,k},\ldots,h_{M,k}]^T\in \mathbb{C}^{M\times 1}$. It is worth emphasizing that the effective channel ${\bf h}_k$ is determined by the joint design of the antenna rotation set ${\bf R}\triangleq\{\mathbf{R}_m\}_{m=1}^{M}$, transmit polarization state set ${\bf V}\triangleq\{\mathbf{v}_m\}_{m=1}^{M}$, and receive polarization state set ${\bf U}\triangleq\{\mathbf{u}_k\}_{k=1}^{K}$.
To characterize the performance limit of the proposed RA architecture, we consider an ideal case in which the effective polarization coupling ${\bf Q}_{m,k,l}{\bf M}_{m,k,l}{\bf P}_{m,k,l}$ is perfectly available.
\subsection{Transmission Model}
\subsubsection{Transmit signal at the BS}
Let $s_k$ denote the transmit signal for user $k$ with $\mathbb{E}[|s_k|^2]=1$, and let ${\bf w}_k \in \mathbb{C}^{M\times 1}$ be the digital beamforming vector for user $k$. The transmitted signal is given by
\begin{align*}
  {\bf x}=\sum_{k=1}^{K}{\bf w}_ks_k.\tag{11} \label{11}
\end{align*}
\subsubsection{Received signal at the UE $k$} The signal received by user $k$ is obtained by combining the signals from the two polarization ports. The received baseband signal $y_k$ can be expressed as 
\begin{align*}
  y_k\!=\!{\bf h}_k^H({\bf V},\!{\bf u}_k,\!{\bf R}) {\bf w}_ks_k\!+\!\sum_{i\neq k}^{K}{\bf h}_k^H({\bf V},\!{\bf u}_k,\!{\bf R}) {\bf w}_is_i\!+\!n_k,\tag{12} \label{12}
\end{align*}
where $n_k \sim \mathcal{CN}(0, \sigma_k^2)$ denotes the additive noise at user $k$. Since the polarization reconfiguration at the user is realized by phase shifters whose noise is negligible compared to the RF chain noise, we only model the RF chain noise in $n_k$~\cite{PRA4}.
Based on \eqref{12}, the SINR of user $k$ is given by
\begin{align*}
  \gamma_k=\frac{|{\bf h}_k^H({\bf V},{\bf u}_k,{\bf R}) {\bf w}_k|^2}{\sum_{i\neq k}^{K}|{\bf h}_k^H({\bf V},{\bf u}_k,{\bf R}) {\bf w}_i|^2+\sigma_k^2},\forall k.\tag{13} \label{13}
\end{align*}
Accordingly, the achievable rate for user $k$ is
\begin{align*}
  R_k=\log_2\left(1+\gamma_k\right),\forall k.\tag{14} \label{14}
\end{align*}
\section{Single-User LoS Scenario}
To quantify the fundamental performance benefits offered by the proposed RA architecture, we consider a simplified single-user LoS scenario with $M=1$, $K=1$, and $L=0$. We focus on the performance improvements brought by 3D rotation and polarization matching, omitting the path loss $\beta$ as it is independent of the RA design.

For the single-user LoS link, let ${\bf f}=[f_x,f_y,f_z]^T\in\mathbb{R}^{3\times 1}$ denote the unit propagation direction vector from the BS to the user. Based on the channel model in \eqref{8}, we define the effective channel gain excluding path loss as 
\begin{align}
  g({\bf v},{\bf u},{\bf R})=G(\varepsilon)\eta({\bf v},{\bf u},{\bf R}),\tag{15} \label{15}
\end{align}
where $G(\epsilon)$ is the directional gain defined in \eqref{5}, and $\eta({\bf v},{\bf u},{\bf R}) = \left| \mathbf{u}^H \mathbf{Q} \mathbf{P} \mathbf{v} \right|^2$ represents the polarization matching efficiency.

To explicitly analyze $\eta$, we utilize the fact that the projection matrix onto the transverse plane is given by
\begin{align}
{\bf Z}{\bf Z}^T={\bf I}_3-{\bf f}{\bf f}^T. \tag{16}\label{16}
\end{align}

Let ${\bf E}=[{\bf e}_h,{\bf e}_v]\in\mathbb{R}^{3\times 2}$ denote the two co-located orthogonal polarization ports with
${\bf e}_h=[1,0,0]^T$ and ${\bf e}_v=[0,1,0]^T$. 
Throughout this subsection, we assume vertically polarized transmission, i.e., ${\bf v}=[0,1]^T$ so that ${\bf E}{\bf v}={\bf e}_v$. By substituting the definitions of $\mathbf{P}$ and $\mathbf{Q}$ into \eqref{16}, the polarization combining gain in \eqref{16} can be rewritten as
\begin{align}
\eta({\bf u},{\bf R})
&= \left|{\bf u}^H({\bf E}^T{\bf Z})({\bf Z}^T{\bf R}{\bf E}{\bf v})\right|^2 \notag\\
&=\left|{\bf u}^H{\bf E}^T({\bf I}_3-{\bf f}{\bf f}^T){\bf R}{\bf e}_v\right|^2, \tag{17}\label{17}
\end{align}
where the second equality follows from \eqref{16}. Considering the phase-only control architecture at the user, we impose a unit-modulus constraint on the combining vector, i.e., $|u_j|=1$ for $j=1,2$. By aligning the phases of ${\bf u}$ to achieve polarization state matching, we obtain
\begin{align}
\eta^\star
\!=\!\!\! \max_{\{|u_j|=1\}}\!\!\eta({\bf u},{\bf R})
\!=\!\left(\!\sum_{j=1}^{2}\left|\left[{\bf E}^T({\bf I}_3-{\bf f}{\bf f}^T){\bf R}{\bf e}_v\right]_j\right|\right)^2\!. \tag{18}\label{18}
\end{align}
Next, we analyze the performance enhancement offered by the proposed RA design relative to the conventional fixed scheme.
\subsubsection*{1) RA with fixed rotation}
When 3D rotation is disabled, we have ${\bf R}={\bf I}_3$. Substituting this into \eqref{18} yields the closed-form combining gain
\begin{align}
\eta_{\rm fix}^\star
&=\left(\sum_{j=1}^{2}\left|\left[{\bf E}^T({\bf I}_3-{\bf f}{\bf f}^T){\bf e}_v\right]_j\right|\right)^2 \notag\\
&=\Big(|f_x f_y|+|1-f_y^2|\Big)^2, \tag{19}\label{19}
\end{align}
which reflects the projection loss since the fixed antenna plane is generally not perpendicular to the propagation direction ${\bf f}$~\cite{nature}. Consequently, the corresponding effective gain is
$g_{\rm fix}^\star = G(\varepsilon_{\rm fix})\,\eta_{\rm fix}^\star$, 
where $\varepsilon_{\rm fix}=\arccos({\bf e}_z^T{\bf f})$ is the misalignment angle under the fixed boresight direction ${\bf e}_z$.

\begin{figure}[t!]
\centering

\subfloat[RA with fixed rotation\label{figLoSa}]{
    \includegraphics[width=0.45\textwidth]{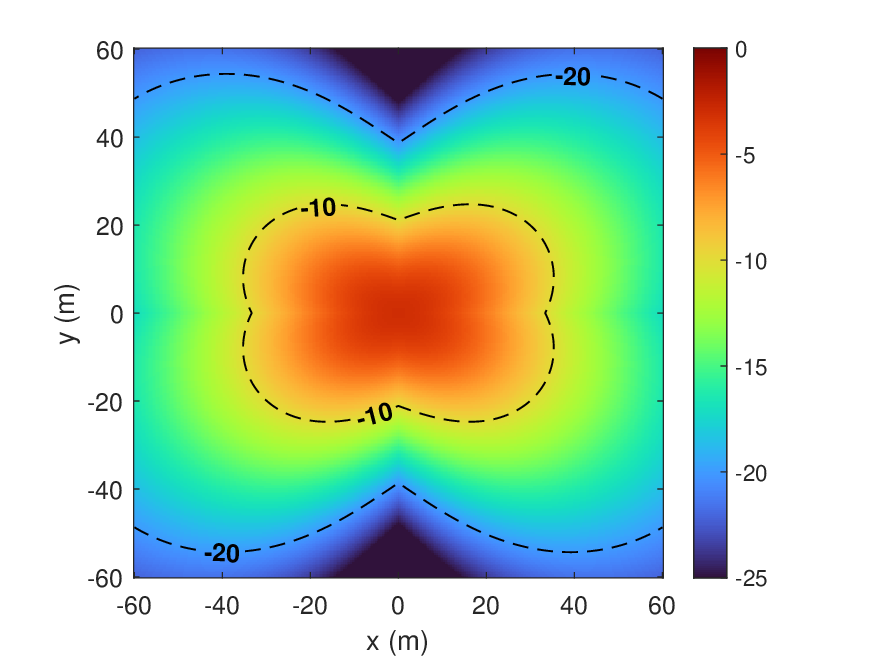}
}\\[0.6em]

\subfloat[RA with optimized rotation\label{figLoSb}]{
    \includegraphics[width=0.45\textwidth]{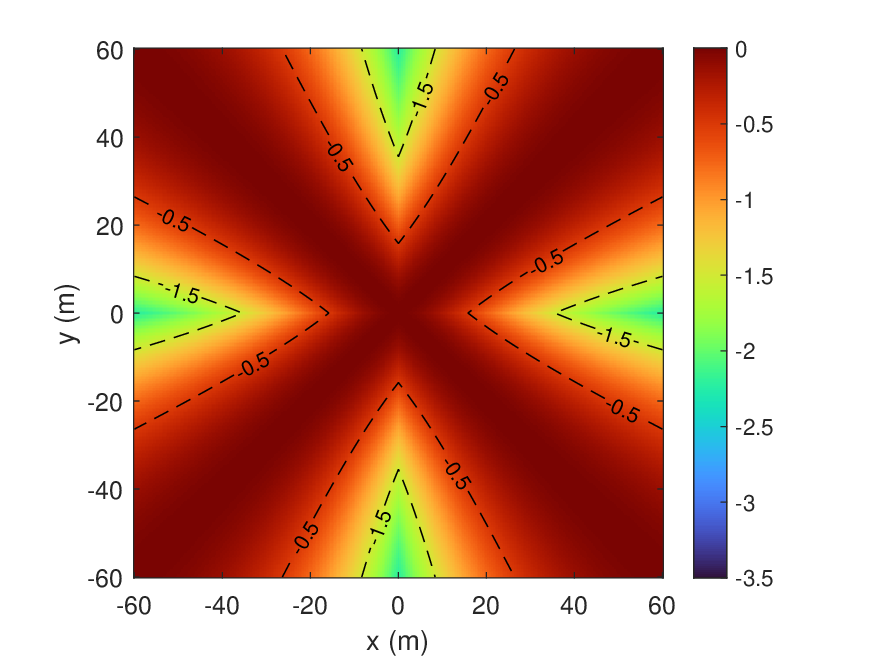}
}

\caption{Spatial distribution of the effective gain}
\label{figureLoS}
\vspace{-1em}
\end{figure}
\subsubsection*{2) RA with optimized rotation}
When rotation is enabled, the matrix ${\bf R}\in \mathrm{SO}(3)$ acts as an optimization variable. 
We first steer the antenna boresight toward the user direction by enforcing ${\bf R}{\bf e}_z={\bf f}$, which ensures $\varepsilon=0$ and maximizes the directional gain to $G(0)=G_0$. Moreover, since ${\bf e}_v\perp{\bf e}_z$, the rotated polarization direction ${\bf r}_{m,2}={\bf R}{\bf e}_v$ satisfies ${\bf r}_{m,2}\perp{\bf f}$. Consequently, the projection term simplifies to
\begin{align}
({\bf I}_3-{\bf f}{\bf f}^T){\bf s}={\bf s},\tag{20}
\label{eq:1}
\end{align}
 which avoids the projection loss at the transmitter side. The optimization problem in \eqref{18} then simplifies to optimizing the direction of $\mathbf{r}_{m,2}$ within the plane orthogonal to ${\bf f}$.
\begin{align}
\eta_{\rm rot}^\star
=\max_{\substack{ \|{\bf r}_{m,2}\|=1,\,{\bf r}_{m,2}\perp{\bf f}}}\left(|[{\bf r}_{m,2}]_1|+|[{\bf r}_{m,2}]_2|\right)^2. \tag{21}\label{21}
\end{align}
A closed-form optimal value of \eqref{21} is
\begin{align}
\eta_{\rm rot}^\star
=2-\min\Big\{(f_x+f_y)^2,\ (f_x-f_y)^2\Big\}. \tag{22}\label{22}
\end{align}
The optimal ${\bf R}^\star$ is defined by aligning ${\bf r}_{m,3}$ with ${\bf f}$ and choosing ${\bf r}_{m,2}$ as the normalized projection of $[1, \pm 1, 0]^T$ onto the null space of ${\bf f}$. The first column is then determined by ${\bf r}_{m,1}={\bf r}_{m,2}\times{\bf r}_{m,3}$.
Finally, the optimal effective gain for the RA architecture is
$g_{\rm rot}^\star = G_0\,\eta_{\rm rot}^\star$.
The theoretical gain ratio between the rotatable and fixed cases can be decomposed as
\begin{align}
\frac{g_{\rm rot}^{\star}}{g_{\rm fix}^{\star}}
=
{\frac{1}{\cos^{2p}(\varepsilon_{\text{fix}})}}
\cdot
{\frac{\eta_{\rm rot}^{\star}}{\eta_{\rm fix}^{\star}}},\tag{23}
\label{23}
\end{align}
which reveals how the 3D rotation contribute to performance from the following two aspects
\begin{itemize}
    \item \textbf{Directional gain} (${{1}/{\cos^{2p}(\varepsilon_{\text{fix}})}}$): This factor originates from {boresight steering}. By adjusting the azimuth and elevation angles to steer the boresight toward the user, the RA mitigates the gain loss caused by angular misalignment under a narrow beam radiation pattern.
    \item \textbf{Polarization matching gain} (${{\eta_{\rm rot}^{\star}}/{\eta_{\rm fix}^{\star}}}$): The second term reflects the gain from 3D rotation via improved polarization direction alignment. Both the fixed and rotatable schemes achieve polarization state matching by aligning the combining phases according to \eqref{18}.
    However, compared with the fixed case, 3D rotation improves the polarization efficiency in two specific ways. First, it steers the transmit polarization to be perpendicular to the propagation direction, thereby avoiding the projection loss based on \eqref{eq:1}. Second, by rolling around the boresight, the RA further optimizes the transmit polarization direction to better match the receive combiner as derived in \eqref{21}.
    For the considered dual-polarized receiver with phase-only combining, ignoring this roll adjustment incurs up to a $3$ dB loss in received power relative to the best roll alignment. By contrast, for single-polarized receivers, polarization mismatch can be far more detrimental and may severely attenuate the received signal \cite{balanis2016antenna}.
\end{itemize}

To visualize the spatial distribution of this performance gap, we conduct a simulation over a plane at $z=30$ m. The BS is located at the origin $(0,0,0)$ m with the fixed antenna boresight aligned with the $z$-axis. The directivity factor is set to $p=2$. The simulation results are normalized by the theoretical maximum $g_{\max}=2G_0$ and presented in Fig. \ref{figureLoS}. Fig. \ref{figureLoS}(a) shows that the fixed antenna suffers from severe gain degradation in the off-boresight regions, where the loss exceeds $20$ dB. This severe performance loss is primarily caused by the beam misalignment attributed to the narrow beam of the directional antenna, which is further aggravated by polarization mismatch. In contrast, Fig. \ref{figureLoS}(b) shows that the proposed RA achieves uniform high-gain coverage across the entire plane.\looseness-1

To quantitatively verify this, we examine a direction $\mathbf{f}=[\sqrt{3/8}, \sqrt{3/8}, 1/2]^T$, corresponding to a misalignment angle of $\varepsilon_{\rm fix}=60^\circ$. In this case, the fixed scheme yields an effective gain of only $g_{\rm fix}^\star = G_0/16$. In contrast, the proposed RA recovers the full directional gain and doubles the polarization efficiency, i.e., $g_{\rm rot}^\star = 2G_0$. This results in a total gain improvement of roughly $15.1 $dB, confirming that the proposed design not only compensates for the $12 $ dB misalignment loss but also provides an additional $3$ dB polarization gain.

\section{General Multi-user Multipath Scenario}
The analysis in Section \uppercase\expandafter{\romannumeral3} validated the fundamental gains arising from the spatial and polarization DoFs. However, extending these insights to general multi-user multipath scenarios presents significant challenges. In these scenarios, the channel induces severe depolarization, and the optimal joint design of rotation and polarization is no longer solely dictated by the LoS direction but is also constrained by inter-user interference and physical rotation angle limits.
Under these challenges, simple boresight alignment or optimizing the rotation matrix in isolation is insufficient. Thus, we formulate a joint optimization problem aimed at minimizing the transmit power by optimizing the digital beamforming, rotation, and polarization designs, subject to the rate requirements of users.\looseness-1
\begin{align}
\mathbf{P}1: \quad & \underset{\mathbf{W}, \mathbf{V}, \mathbf{U}, \mathbf{R}}{\min} \quad \sum_{k=1}^K \|\mathbf{w}_k\|^2 \tag{24} \label{24} \\
\text{s.t.} \quad &  R_k \ge \bar R_{ k}, \; \forall k , \tag{24a} \label{24a} \\
& |{u}_{k,j}| = 1, \; j=1,2,\forall k,\tag{24b} \label{24b} \\
& \|\mathbf{v}_m\|_2 = 1, \; \forall m , \tag{24c} \label{24c}\\
&  \mathbf{R}_m^T\mathbf{R}_m={\bf{I}}, \; \forall m , \tag{24d} \label{24d}\\
& \det{(\mathbf{R}_m)}=1, \; \forall m, \tag{24e} \label{24e}\\
&  0\le\arccos\left( \mathbf{r}_{m,3}^T {\bf e}_z \right) \le \theta_{\max}, \;\forall m, \tag{24f} \label{24f}
\end{align}
where ${\bf W} \triangleq\{\mathbf{w}_k\}_{k=1}^{K}$ collects the digital beamforming vectors at the BS. 
Constraint \eqref{24a} enforces the minimum rate requirement $\bar{R}_k$ for each user $k$. Constraint \eqref{24b} imposes the unit-modulus constraints on the entries of ${\bf u}_k=[u_{k,1},u_{k,2}]^T$, consistent with the phase-only control architecture at the user receivers. Constraint \eqref{24c} represents the power normalization for the polarization state vector ${\bf v}_m$ at the BS. Constraint \eqref{24d} and \eqref{24e} restrict $\mathbf{R}_m$ to a rigid-body rotation, ensuring that the antenna orientation is adjusted without changing its physical structure. Finally, constraint \eqref{24f} confines the boresight deviation angle within $[0,\theta_{\max}]$, which accounts for practical hardware limitations and also helps mitigate antenna coupling.
\section{Proposed Solutions}
In this section, we develop an algorithm to solve problem $\mathbf{P}1$. Given the strong coupling among variables, problem $\mathbf{P}1$ is highly non-convex and intractable to solve directly. 
To address this, we propose an AO framework employing optimization strategies tailored to the specific constraints of each variable. Specifically, for the digital beamforming $\mathbf{W}$, we employ a combination of SDP and DC method to handle the non-convex quadratic terms. Conversely, for the rotation $\mathbf{R}$ and polarization variables $\{\mathbf{V}, \mathbf{U}\}$, which are strictly confined to the $SO(3)$ group, complex unit sphere, and the complex circle manifold, respectively, we utilize the RCG method to efficiently perform updates directly on their respective manifolds. Finally, we analyze the convergence behavior and the computational complexity of the proposed algorithm.
 \subsection{Optimization of the Digital Beamforming at BS}
In this subsection, we optimize the digital beamforming vectors while keeping the rotation and polarization variables fixed. 
With ${\bf R}$, ${\bf V}$ and ${\bf U}$ given, the original problem reduces to the digital beamforming subproblem $\mathbf{P}2$:
\begin{align}
\mathbf{P}2: \quad & \underset{\mathbf{W}}{\min} \quad \sum_{k=1}^K \|\mathbf{w}_k\|^2 \notag \\
\text{s.t.} \quad &  \eqref{24a}.\notag
\end{align}
The above problem remains non-convex due to the quadratic terms coupled in the SINR constraints. To address this, we reformulate the problem using a SDP approach combined with DC techniques. Specifically, we introduce the PSD matrices ${\bf D}_k={\bf w}_k{\bf w}_k^H,\forall k$ to linearize the quadratic terms. Consequently, the subproblem is reformulated as
\begin{align*}
\mathbf{P}2.1: &\underset{\{\mathbf{D}_k\}_{k=1}^K}{\min}  \sum_{k=1}^K \text{tr}({\bf D}_k) \tag{25} \label{25} \\
\text{s.t.} \quad &  {\bf h}_k^H{\bf D}_k{\bf h}_k\ge \iota_k\left(\sum_{i\neq k}^K{\bf h}_k^H{\bf D}_i{\bf h}_k+\sigma_k^2\right) ,\forall k,\tag{25a} \label{25a}\\
& \text{rank}({\bf D_k})=1,\forall k,\tag{25b} \label{25b}
\end{align*}
where $\iota_k=2^{\bar{R}_k}-1$.
The reformulation linearizes the quadratic terms but introduces the non-convex rank-one constraint \eqref{25b}. To address this, we leverage the fact that the rank-one constraint is equivalent to the difference of two convex functions
\begin{align}
\text{rank}(\mathbf{D}_k)=1 \iff \mathrm{tr}(\mathbf{D}_k) - \|\mathbf{D}_k\|_2 = 0, \tag{26} \label{26}
\end{align}
where $\|\cdot\|_2$ denotes the spectral norm. To enforce the rank-one constraint \eqref{25}, we directly add the penalty term $\lambda\sum_{k=1}^K (\mathrm{tr}({\bf D}_k)-\|{\bf D}_k\|_2)$ to the objective function. Since $-\|{\bf D}_k\|_2$ is concave, we linearize it using the first-order Taylor expansion at the previous iterate ${\bf D}_{k,t-1}$, yielding the upper bound $-\|{\bf D}_k\|_2
\ \le\ 
-\|{\bf D}_{k,t-1}\|_2
-\left\langle {\bf d}_{k,t-1}{\bf d}_{k,t-1}^H,\ {\bf D}_k-{\bf D}_{k,t-1}\right\rangle$,
where $\langle{\bf X},{\bf Y}\rangle= \Re\{\mathrm{tr}({\bf X}^H{\bf Y})\}$, ${\bf d}_{k,t-1}$ denotes the eigenvector corresponding to the largest eigenvalue of ${\bf D}_{k,t-1}$~\cite{11012663}. By substituting this approximation into the objective function and omitting the constant terms independent of ${\bf D}_k$, the subproblem is formulated as
\begin{align}
\mathbf{P}2.2
\min_{\{{\bf D}_k\}_{k=1}^K} 
& \!\sum_{k=1}^K \!\mathrm{tr}({\bf D}_k)
\!+\!\lambda_0 \!\!\sum_{k=1}^K \!\!\left(
\mathrm{tr}({\bf D}_k)
\!-\!\left\langle \!{\bf d}_{k,t-\!1}{\bf d}_{k,t-\!1}^H,\! {\bf D}_k \right\rangle
\!\right) \notag\\
\text{s.t.}\quad 
& \eqref{25a}. \notag
\end{align}
The reformulated subproblem $\mathbf{P}2.2$ is convex and can be efficiently solved by CVX tools \cite{grant2008cvx}.
\subsection{Optimization of the Rotation Matrices at BS}
In this subsection, we optimize the rotation matrices ${\bf R} \triangleq\{{\bf R}_m\}_{m=1}^M$ with the digital beamforming and polarization variables fixed. 
Observe that the rotation matrices ${\bf R}$ do not appear in the objective function; instead, they directly determine the quality of the effective channels $\mathbf{h}_k(\mathbf{R})$. By optimizing $\mathbf{R}$, we maximize the channel gain, which improves the feasibility margin of the SINR constraints, allowing for reduced transmit power.
Therefore, instead of directly minimizing the transmit power, we update ${\bf R}$ by improving the worst-user link quality, i.e., we consider the following max-min SINR subproblem
\begin{align*}
  \mathbf{P}3
\max_{\{{\bf R}_m\}_{m=1}^M} &\min 
 \{\gamma_1,\ldots,\gamma_K\} \\
\text{s.t.}\quad 
& \eqref{24a},\eqref{24d},\eqref{24e},\eqref{24f}. \notag
\end{align*}
The constraints \eqref{24d} and \eqref{24e} imply that the rotation matrices lie on the product manifold $\mathcal{M}_R \triangleq \mathrm{SO}(3)^M$. This smooth Riemannian structure enables the application of the RCG method.
To handle the remaining constraints \eqref{24a} and \eqref{24f} within the manifold framework, we incorporate them into the objective function via quadratic penalty terms. Consequently, problem $\mathbf{P}3$ is reformulated as minimizing the following penalty-based objective function over $\mathcal{M}_R$ 
\begin{align}
\widetilde{\mathcal{J}}_R({\bf R})
=
&\max\{-\gamma_1({\bf R}),\ldots,-\gamma_K({\bf R})\} \notag \\
&+\lambda_1 \widetilde{P}_{\rm angle}({\bf R})
+\lambda_2 \widetilde{P}_{\rm thresh}({\bf R}),
\tag{27} \label{27}
\end{align}
where $\lambda_1$ and $\lambda_2$ denote the penalty parameters. The terms $\widetilde{P}_{\rm angle}(\cdot)$ and $\widetilde{P}_{\rm thresh}(\cdot)$ quantify the violations of the rotation constraint \eqref{24f} and the SINR requirements \eqref{24a}, respectively. Specifically, utilizing the equivalent condition ${\bf r}_{m,3}^T{\bf e}_z\ge \cos\theta_{\max}$ for the angle constraint, we define these penalty terms as
\begin{align}
\tilde P_{\rm angle}({\bf R})
&=\sum_{m=1}^{M}\big(\max\{0,\ \cos\theta_{\max}-{\bf r}_{m,3}^T{\bf e}_z\}\big)^2, \tag{28a}\label{28a}\\
\tilde P_{\rm thresh}({\bf R})
&=\sum_{k=1}^{K}\big(\max\{0,\ \iota_k-\gamma_k({\bf R})\}\big)^2.
\tag{28b}\label{28b}
\end{align}

However, the objective function $\widetilde{\mathcal{J}}_R({\bf R})$ is non-differentiable due to the $\max(\cdot)$ operators. To address this, we apply the Log-Sum-Exp function to approximate the term $\max\{-\gamma_1({\bf R}),\ldots, -\gamma_K({\bf R})\}$ as
\begin{align}
f_{\rm obj}({\bf R})
=\frac{1}{\mu}\ln\!\left(\sum_{k=1}^{K}\exp\!\left(-\mu\,\gamma_k({\bf R})\right)\right),
\tag{29} \label{29}
\end{align}
where $\mu > 0$ is a smoothing parameter. Minimizing the smooth upper bound effectively maximizes the minimum SINR. For the penalty terms, we employ the Softplus function $\mathrm{sp}_\alpha(x)\!=\!\frac{1}{\alpha}\ln(1+\exp(\alpha x))$ to approximate $\max\{0,x\}$, yielding\looseness-1
\begin{align}
P_{\rm angle}({\bf R})
&=\sum_{m=1}^{M}\big(\mathrm{sp}_\alpha(\cos\theta_{\max}-{\bf r}_{m,3}^T{\bf e}_z)\big)^2,\tag{30a} \label{30a}\\
P_{\rm thresh}({\bf R})
&=\sum_{k=1}^{K}\big(\mathrm{sp}_\alpha(\iota_k-\gamma_k({\bf R}))\big)^2.
\tag{30b}\label{30b}
\end{align}

By combining the smooth surrogates derived in \eqref{29}, \eqref{30a} and \eqref{30b}, we construct the continuously differentiable objective function $\mathcal{J}_R({\bf R})$ on the manifold $\mathcal{M}_R$ as:
\begin{align} 
\mathcal{J}_R({\bf R})
=
f_{\rm obj}({\bf R})+\lambda_1 P_{\rm angle}({\bf R})+\lambda_2 P_{\rm thresh}({\bf R}),
\tag{31} \label{31}
\end{align}
To enforce the constraints, the penalty parameters $\lambda_1$ and $\lambda_2$ are dynamically increased if the corresponding constraints are not satisfied.

The smoothness of $\mathcal{J}_R({\bf R})$ enables the computation of Euclidean gradients and the application of the RCG algorithm. Consequently, we proceed to solve the problem $\mathbf{P}3.1$ via the RCG algorithm
\begin{align*}
  \mathbf{P}3.1
\min_{{\bf R}\in\mathcal{M}_R}  \mathcal{J}_R({\bf R}),
\end{align*}
The main steps of the RCG algorithm on $\mathrm{SO}(3)^M$ are summarized as follows
\begin{itemize}
\item \textbf{Riemannian Gradient:} 
Let ${\bf G}_m = \nabla_{{\bf R}_m}\mathcal{J}_R({\bf R}) \in \mathbb{R}^{3\times 3}$ denote the Euclidean gradient of $\mathcal{J}_R({\bf R})$ at ${\bf R}_m$. The detailed derivation of ${\bf G}_m$ is provided in Appendix A.
The Riemannian gradient on the tangent space of $\mathrm{SO}(3)$ is obtained by the orthogonal projection of ${\bf G}_m$:
\begin{align}
\mathrm{grad}_{{\bf R}_m}\mathcal{J}_R
= {\bf G}_m - {\bf R}_m\,\mathrm{sym}\left({\bf R}_m^T{\bf G}_m\right),
\tag{32} \label{32}
\end{align}
where $\mathrm{sym}({\bf A})= \tfrac{1}{2}({\bf A}+{\bf A}^T)$.
The full Riemannian gradient is given by $\mathrm{grad}_{\bf R}\mathcal{J}_R=\{\mathrm{grad}_{{\bf R}_m}\mathcal{J}_R\}_{m=1}^M$.

\item \textbf{Update of search direction.}
At iteration $t$, 
the conjugate search direction is updated as
\begin{align}
\boldsymbol{\xi}_m^{(t)}
=
-\mathrm{grad}_{{\bf R}_m}\!\mathcal{J}_R({\bf R}^{(t)}\!)
\!+\!\beta^{(t)}\!\mathcal{T}_{{\bf R}_m^{(t-1)}\!\rightarrow{\bf R}_m^{(t)}}\left(\!\boldsymbol{\xi}_m^{(t-1)}\!\right)\!,
\tag{33} \label{33}
\end{align}
where $\mathcal{T}(\cdot)$ denotes a vector transport between tangent spaces on $\mathrm{SO}(3)$ and $\beta^{(t)}$ is computed by the Polak--Ribi\`ere rule \cite{yu2019miso}.

\item \textbf{Stepsize Selection:}
The stepsize $\alpha^{(t)}$ is determined via an Armijo backtracking line search on $\mathcal{M}_R$ along the search direction $\{\boldsymbol{\xi}_m^{(t)}\}_{m=1}^M$.

\item \textbf{Retraction.}
To maintain feasibility on $\mathrm{SO}(3)$, we retract the tangent update back to the manifold.
Given ${\bf Y}_m^{(t)}={\bf R}_m^{(t)}+\alpha^{(t)}\boldsymbol{\xi}_m^{(t)}$, we adopt the polar retraction
\begin{align}
{\bf R}_m^{(t+1)}
= {\bf A}\,\mathrm{diag}\!\big(1,1,\det({\bf A}{\bf B}^T)\big)\,{\bf B}^T,\tag{34} \label{34}
\end{align}
where ${\bf Y}_m^{(t)}={\bf A}{\bf \Sigma}{\bf B}^T$ is the singular value decomposition (SVD) of ${\bf Y}_m^{(t)}$.
\end{itemize}

\subsection{Optimization of the Transmit Polarization Design at BS}
In this subsection, we optimize the transmit polarization vectors ${\bf V}\triangleq\{\mathbf{v}_m\}_{m=1}^{M}$ with other variables fixed. 
Similar to the optimization of ${\bf R}$, we update ${\bf V}$ by solving the following max-min SINR subproblem:
\begin{align}
\mathbf{P}4:\quad
\max_{\{{\bf v}_m\}_{m=1}^M}\ \min \{\gamma_1,\ldots,\gamma_K\} 
\quad
\text{s.t.}\ 
\eqref{24a},\eqref{24c},
\notag
\end{align}
The constraints in \eqref{24c} restrict the feasible set to a manifold $\mathcal{M}_V$, defined as the product of $M$ complex unit spheres:
\begin{align}
\mathcal{M}_V  \triangleq \{{\bf V}\mid \|{\bf v}_m\|_2=1,\ \forall m\}.\tag{35}
\label{35}
\end{align}
Since $\mathcal{M}_V$ is a smooth manifold, we can also apply the RCG algorithm. To handle the inequality constraints \eqref{24a}, we incorporate them into the objective via a penalty term. Furthermore, utilizing the same smooth surrogate functions as in the rotation optimization to approximate the non-differentiable components, we obtain the following smooth unconstrained minimization problem on $\mathcal{M}_V$
\begin{align}
\mathbf{P}4.1 \min_{{\bf V}\in\mathcal{M}_V}\ 
\mathcal{J}_V({\bf V})=
&f_{\rm obj}({\bf V})+\lambda_3 P_{\rm thresh}({\bf V}),
\tag{36}\label{36}
\end{align}
where $\lambda_3 > 0$ is the penalty parameter.
We then apply RCG to solve $\mathbf{P}4.1$ on $\mathcal{M}_V$. 
At iteration $t$, let ${\bf G}_m^{(t)} = \nabla_{{\bf v}_m^\ast}\mathcal{J}_V({\bf V})$ denote the Euclidean gradient derived in Appendix B. 
The Riemannian gradient is obtained by projecting ${\bf G}_m^{(t)}$ onto the tangent space
\begin{align}
\mathrm{grad}_{{\bf v}_m}\mathcal{J}_V
=
\left({\bf I}_2-{\bf v}_m{\bf v}_m^H\right)\,{\bf G}_m^{(t)},\; \forall m.
\tag{37}\label{37}
\end{align}
With the Riemannian gradients, the search direction $\boldsymbol{\xi}_m^{(t)}$ is updated following the same conjugate gradient rule and vector transport mechanism as given in \eqref{33}. The stepsize $\alpha^{(t)}$ is determined via an Armijo backtracking line search.
Finally, we retract the updated tangent vectors back to $\mathcal{M}_V$ by normalization:
\begin{align}
{\bf v}_m^{(t+1)}
=
\frac{{\bf v}_m^{(t)}+\alpha^{(t)}\boldsymbol{\xi}_m^{(t)}}
{\big\|{\bf v}_m^{(t)}+\alpha^{(t)}\boldsymbol{\xi}_m^{(t)}\big\|_2},
\; \forall m.
\tag{38}\label{38}
\end{align}

\subsection{Optimization of the Receive Polarization Design at Users}
In this subsection, we optimize the receive polarization combining vectors ${\bf U}\triangleq\{\mathbf{u}_k\}_{k=1}^{K}$ with other variables fixed. 
Similar to the optimization of ${\bf R}$ and ${\bf V}$, we update ${\bf U}$ by solving the following max-min SINR problem to improve the link quality
\begin{align}
\mathbf{P}5:\quad
\max_{\{{\bf u}_k\}_{k=1}^K}\ \min \{\gamma_1,\ldots,\gamma_K\} 
\quad
\text{s.t.}\ 
\eqref{24a},\eqref{24b},
\notag
\end{align}
The constraints \eqref{24b} restrict the feasible set to the complex circle manifold $\mathcal{M}_U$
\begin{align}
\mathcal{M}_U
 \triangleq\Big\{{\bf U}\ \big|\ |[{\bf u}_k]_j|=1,\ \forall k,\ j=1,2\Big\}.
\tag{39}\label{39}
\end{align}
Adopting the same smoothing strategies as in previous subsections, we formulated the smooth problem on $\mathcal{M}_U$
\begin{align}
\mathbf{P}5.1: \quad
\min_{{\bf U}\in\mathcal{M}_U} \ 
\mathcal{J}_U({\bf U})
=
f_{\rm obj}({\bf U})+\lambda_4 P_{\rm thresh}({\bf U}).
\tag{40}\label{40}
\end{align}
We apply the RCG algorithm to solve $\mathbf{P}5.1$. Let ${\bf G}_k^{(t)}=\nabla_{{\bf u}_k^\ast}\mathcal{J}_U({\bf U})$ denote the Euclidean gradient (see Appendix C).
The Riemannian gradient is obtained by projecting ${\bf G}_k^{(t)}$ onto the tangent space via element-wise operations:
\begin{align}
\mathrm{grad}_{{\bf u}_k}\mathcal{J}_U
=
{\bf G}_k^{(t)}-\Re\!\left\{{\bf G}_k^{(t)}\odot({\bf u}_k^{(t)})^\ast\right\}\odot{\bf u}_k^{(t)},
\tag{41}\label{41}
\end{align}
where $\odot$ denotes the Hadamard product.
The search direction $\boldsymbol{\xi}_k^{(t)}$ is updated following the same rule as in \eqref{33}.
Finally, the updated points are retracted to $\mathcal{M}_U$ by 
\begin{align}
{\bf u}_k^{(t+1)}
=
\exp\!\Big(j\,\angle\!\big({\bf u}_k^{(t)}+\alpha^{(t)}\boldsymbol{\xi}_k^{(t)}\big)\Big),
\; \forall k.
\tag{42}\label{42}
\end{align}

In summary, the overall algorithm to solve the
problem $\mathbf{P}1$ is summarized in Algorithm \ref{alg:AO_RCG}.
\subsection{Algorithm Discussion}
\subsubsection{Convergence Analysis}
Let $P^{(i-1)}= \sum_{k=1}^{K}\|{\bf w}_k^{(i-1)}\|^2$ denote the transmit power after updating ${\bf W}$ in the $(i-1)$-th AO iteration. To ensure the convergence of the AO algorithm, it is essential to guarantee that the transmit power is non-increasing across the AO iterations. At the $(i-1)$-th AO iteration, we have a feasible solution
$\big({\bf W}^{(i-1)},{\bf V}^{(i-1)},{\bf U}^{(i-1)},{\bf R}^{(i-1)}\big)$
that satisfies all the constraints. In the subsequent updates of ${\bf R}$, ${\bf V}$, and ${\bf U}$, we do not change ${\bf W}$, hence the transmit power remains unchanged, i.e., $P^{(i-1)}$ throughout the updates of ${\bf R}$, ${\bf V}$, and ${\bf U}$.
Moreover, the updates of ${\bf R}$, ${\bf V}$, and ${\bf U}$ are designed to improve the worst-user link quality. 
Therefore, once the rate constraints are satisfied during these updates, the resulting effective channels yield no smaller SINRs under the fixed beamforming ${\bf W}^{(i-1)}$. Thus, the feasible set for the power minimization problem is effectively expanded to further reduce the transmit power, resulting in
$P\!\left(\!{\bf W}^{(i)}\!,\!{\bf V}^{(i)}\!,\!{\bf U}^{(i)}\!,\!{\bf R}^{(i)}\!\right)
\!\le\!\!
P\!\left(\!{\bf W}^{(i-1)}\!,\!{\bf V}^{(i)}\!,\!{\bf U}^{(i)}\!,\!{\bf R}^{(i)}\!\right)=P^{(i-1)}$. Consequently, $P^{(i)}\le P^{(i-1)}$, i.e., the transmit power is non-increasing over the AO iterations.
Therefore, the transmit power sequence ${P^{(i)}}$ is guaranteed to converge. Moreover, the proposed AO algorithm converges to a local solution of the original problem.

\subsubsection{Computational Complexity}
We analyze the computational complexity in terms of the outer AO iterations and the inner iterations introduced by the penalty updates.
Let $I$ be the number of outer AO iterations. For the four subproblems, let $I_W$, $I_R$, $I_V$, and $I_U$ denote the numbers of penalty-update iterations required to satisfy the constraints, respectively. Using an interior-point method, the computational complexity of solving $\mathbf{P}2.1$ is
${S}_1=\mathcal{O}\left((MK)^{4.5}\log(1/\varsigma)\right)$,
where $\varsigma$ denotes the solution accuracy~\cite{5447068}. Moreover, the computational cost of the RCG method is dominated by the corresponding gradients. 
Using a matrix implementation, the cost scales as $\mathcal{O}(MK^2+MKL)$ every RCG iteration. The term $MKL$ arises because the rotation variables affect the channel nonlinearly, so the gradient evaluation requires reconstructing the effective channel from the $L$ multipath components in every iteration.
Therefore, the complexity of solving the subproblem on ${\bf R}$ by RCG is
$
S_2=\mathcal{O}\big(I_{\mathrm{RCG,R}}(MK^2+MKL\big),
$
where $I_{\mathrm{RCG,R}}$ denotes the number of RCG iterations for the subproblem on ${\bf R}$.
In contrast, for ${\bf V}$ and ${\bf U}$, the gradients depend linearly on the channel. Thus, the aggregated channel can be pre-calculated and stored once.
Consequently, the complexity for solving the ${\bf V}$-subproblem is $
S_3=\mathcal{O}\big(I_{\mathrm{RCG,V}}MK^2\big),
$
where $I_{\mathrm{RCG,V}}$ denotes the number of RCG iterations on $\bf V$. Similarly, for the subproblem on ${\bf U}$, the complexity is $S_4=\mathcal{O}\big(I_{\mathrm{RCG,U}}MK^2\big)$. Thus, the overall computational complexity of the proposed algorithm is $I\left(I_WS_1+I_RS_2+I_VS_3+I_US_4\right)$.
\begin{algorithm}[tb]
  \caption{Proposed AO Algorithm for Problem $\mathbf{P}1$}
  \label{alg:AO_RCG}
  \begin{algorithmic}[1] 
    \Require
    Set iteration index $i = 0$.
    Initialize variables $\mathbf{W}^{(0)}$, $\mathbf{R}^{(0)}$, $\mathbf{V}^{(0)}$, $\mathbf{U}^{(0)}$.
    Initialize penalty parameters $\lambda_j, \forall j$, and set scaling factor $\tau > 1$.
    \Ensure
      Obtain the optimized designs $\mathbf{W}^\star$, $\mathbf{R}^\star$, $\mathbf{V}^\star$, and $\mathbf{U}^\star$.

    \Repeat

\State Given $\mathbf{W}^{(i)}$, $\mathbf{V}^{(i)}$, and $\mathbf{U}^{(i)}$. Set inner iteration \State index $t=1$ and penalty parameters $\lambda_1$,$\lambda_2$.
\While{Constraints \eqref{24a} or \eqref{24f} are not satisfied}
    \State Update $\mathbf{R}_t^{(i+1)}$ by solving $\mathbf{P}3.1$ via RCG method.
    \State Update $\lambda_1 \leftarrow \tau\lambda_1$ if \eqref{24f} is violated; update $\lambda_2 \leftarrow $ \State  $\tau\lambda_2$ if \eqref{24a} is violated. Set $t \leftarrow t+1$.
\EndWhile

\State Given $\mathbf{W}^{(i)}$, $\mathbf{R}^{(i+1)}$, and $\mathbf{U}^{(i)}$. Set inner iteration \State index $t=1$ and penalty parameters $\lambda_3$.
\While{Constraint \eqref{24a} is not satisfied}
    \State Update $\mathbf{V}_t^{(i+1)}$ by solving $\mathbf{P}4.1$ via RCG method.
    \State Update $\lambda_3 \leftarrow \tau\lambda_3$ and $t \leftarrow t+1$.
\EndWhile
      \State Given $\mathbf{W}^{(i)}$, $\mathbf{R}^{(i+1)}$, and $\mathbf{V}^{(i+1)}$. Set inner iteration \State index $t=1$ and penalty parameters $\lambda_4$.
\While{Constraint \eqref{24a} is not satisfied}
    \State Update $\mathbf{U}_t^{(i+1)}$ by solving $\mathbf{P}5.1$ via RCG method.
    \State Update $\lambda_4 \leftarrow \tau\lambda_4$ and $t \leftarrow t+1$.
\EndWhile
\State Given $\mathbf{R}^{(i+1)}$, $\mathbf{V}^{(i+1)}$, and $\mathbf{U}^{(i+1)}$. Set inner iteration  \State index $t=1$ and penalty parameter $\lambda_0$.
\While{The rank-one constraint \eqref{25b} is not satisfied}
    \State Update $\{\mathbf{D}_{k,t}\}_{k=1}^K$ by solving the problem $\mathbf{P}2.2$.
    \State Update penalty parameter $\lambda_0 \leftarrow \tau\lambda_0$ and $t \leftarrow t+1$.
\EndWhile
\State Recover the digital beamforming ${\bf W}^{(i+1)}$ via SVD.
      \State Update iteration index $i \leftarrow i + 1$.
    \Until{The value of objective function converges.}
  \end{algorithmic}
\end{algorithm}

\section{Simulations} 
In this section, we evaluate the performance of the proposed system. We start by validating the convergence of our algorithms. Numerical results then confirm that in general multi-user multi-path scenarios, the proposed joint optimization of antenna rotation and polarization significantly reduces the transmit power consumption compared to existing schemes.

\begin{table}[t]
    {\caption{Simulation Parameters}\label{table2}}
    
    \centering
    {
    \begin{tabular}{|c| c| c|}
        \hline
        Parameter & Symbol & Value   \\
        \hline
        Carrier frequency &-& $2.4$ GHz\\
        Number of BS antennas & $M$  & $16 \;(4\times 4)$\\
        Number of scatterers & $L$ & $8$ \\
        Number of users & $K$ & $5$\\
        User distance range & - & $[30, 60]$ m \\
        User angle range (Z-axis) & - & $[0^\circ, 70^\circ]$ \\
        Noise power & $\sigma^2$ & $-80$ dBm \\
        Required Rate & $\bar R$ & $1 - 5$ bps/Hz \\
        Directivity factor & $p$ & $0 - 5$ \\
        Co-polarization power split & $\chi$ & $0.9$ (XPD $= 9$)\\
        Maximum rotation angle & $\theta_{{\max}}$ & $\frac{\pi}{5}$\\
        \hline
    \end{tabular}}
    \vspace{-1em}
\end{table}

Unless otherwise specified, the simulation parameters are set as follows.
The carrier frequency is $2.4$ GHz. At the BS side, a UPA with $M=16$ antennas is employed, arranged in a $4\times 4$ grid with half-wavelength spacing. The number of users is set to $K=5$. The downlink channel is generated with one LoS path and $L=8$ NLoS scattering paths.
The noise power at each user is fixed at $\sigma^2 = -80$ dBm~\cite{RAA1}. Regarding the user distribution, the distance between the BS and each user is uniformly generated within the range of $[30, 60]$ m, and the angle of user relative to the BS Z-axis is uniformly distributed in $[0^\circ, 70^\circ]$. For the NLoS path, the co-polarization power fraction is set to $\chi = 0.9$, implying an XPD of $9$.
For the directional antenna, the directivity factor $p$ is varied from $0$ to $5$ to evaluate different beamwidths, and the initial orientation of the antennas is set to be aligned with the Z-axis. Key simulation parameters are summarized in Table \ref{table2}.

To explicitly evaluate the performance gains contributed by the proposed joint design, we compare the Proposed RA Scheme against three benchmark schemes. For fair comparison, all schemes share identical rate constraints and DBF optimization frameworks, differing only in their antenna capabilities. The specific schemes and their abbreviations used in the simulation figures are defined as follows:
\begin{itemize}
  \item \textbf{Proposed RA Scheme (Proposed Rot+Pol RA + DBF):} Here, Rot and Pol denote rotation and polarization, respectively. In this scheme, the BS antennas support 3D rotation, while both the BS and user antennas enable flexible polarization reconfiguration. Consequently, the digital beamforming matrix ${\bf W}$, the rotation matrices $\{{\bf R}_m\}$, and the polarization states (${\bf V}, {\bf U}$) are jointly optimized to minimize the total transmit power.
\item \textbf{Rotation-Only RA Scheme (Rotation-Only RA + DBF):} In this scheme, the BS antennas are rotatable but lack polarization reconfigurability. The digital beamforming ${\bf W}$ and rotation matrices $\{{\bf R}_m\}$ are optimized, while the polarization states are fixed. This baseline isolates the power saving gain provided by rotation.
\item \textbf{Boresight-Only RA Scheme (Boresight-Only RA + DBF):} The BS antennas optimize only their boresight directions $\mathbf{r}_{m,3}$. The orientation of the polarization directions, $\mathbf{r}_{m,1}$ and $\mathbf{r}_{m,2}$, is deterministically coupled to $\mathbf{r}_{m,3}$ assuming zero rotation around the boresight axis.
This benchmark represents conventional rotatable antennas that prioritize only directional gain.

\item \textbf{Conventional Fixed-Antenna Scheme (Fixed UPA + DBF):} This represents the conventional fixed UPA system. Both the rotation and polarization states are fixed. Only the digital beamforming matrix ${\bf W}$ is optimized. This scheme serves as the performance lower bound.
\end{itemize}

\begin{figure}[t!]
  \centering
  \includegraphics[width=0.48\textwidth]{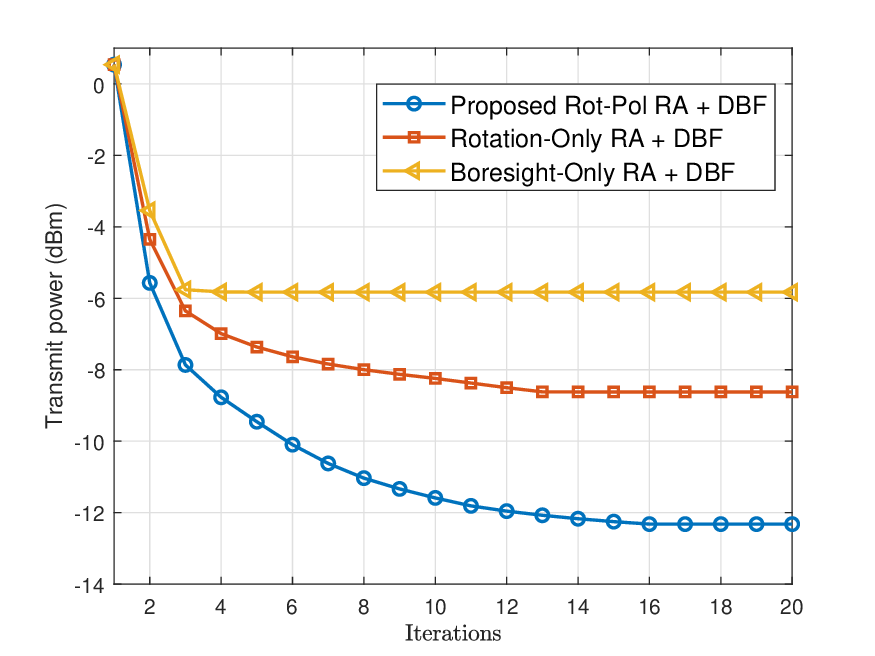}
  \caption{Transmit power versus iteration number}
     \vspace{-0.5em}
  \label{figure2}
\end{figure}
\begin{figure}[t!]
  \centering
  \includegraphics[width=0.48\textwidth]{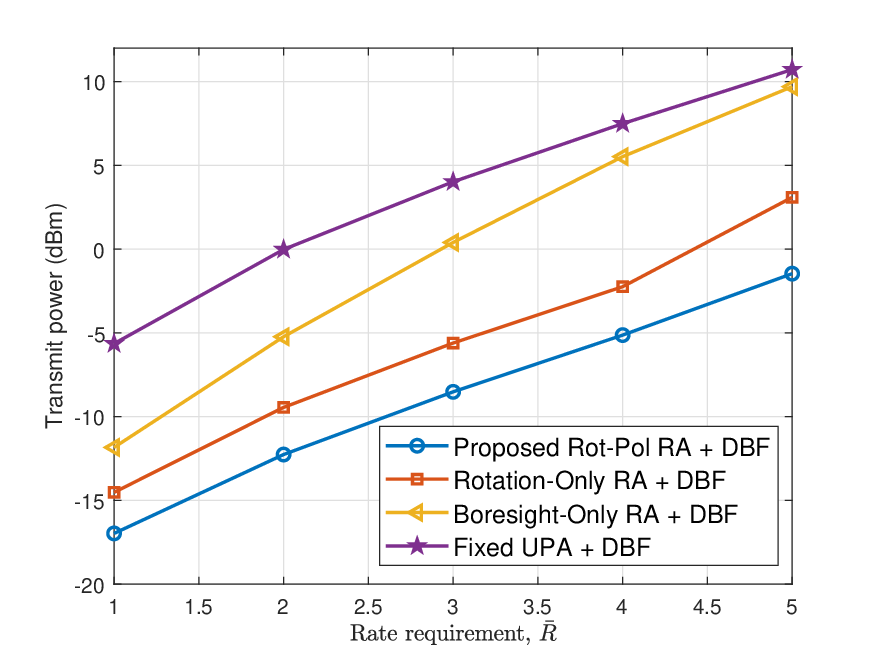}
  \caption{Transmit power versus rate requirement, $\bar R$}
  \label{figure3}
    \vspace{-0.5em}
\end{figure}

Fig.~\ref{figure2} illustrates the convergence behavior of the Proposed Rot-Pol RA + DBF scheme alongside the Rotation-Only and Boresight-Only benchmarks. The simulation is conducted with a rate requirement of $\bar{R} = 2$ bps/Hz and a directivity factor of $p=2$. It can be observed that the total transmit power of all three schemes decreases monotonically over the iterations and stabilizes within approximately $15$ iterations. This rapid convergence behavior validates the efficiency and practicality of the proposed algorithm.

Fig.~\ref{figure3} investigates the total transmit power versus the rate requirement $\bar{R}$, with the directivity factor set to $p=2$. It is observed that the transmit power for all schemes increases monotonically with the growth of the rate requirement. Moreover, the Proposed Rot-Pol RA + DBF scheme consistently outperforms the three benchmark schemes. Specifically, compared to the conventional Fixed UPA + DBF scheme, the proposed method reduces the transmit power by approximately $12$ dB. Moreover, the Rotation-Only RA + DBF scheme consistently outperforms the Boresight-Only RA + DBF benchmark, indicating that optimizing the 3D rotation yields additional benefits beyond boresight steering. Furthermore, the performance gap becomes more pronounced as the rate requirement $\bar{R}$ increases. At higher $\bar{R}$, the SINR constraints become more stringent, so any polarization mismatch in the boresight-only design must be offset by additional transmit power, which enlarges the gap.

\begin{figure}[t!]
  \centering
  \includegraphics[width=0.48\textwidth]{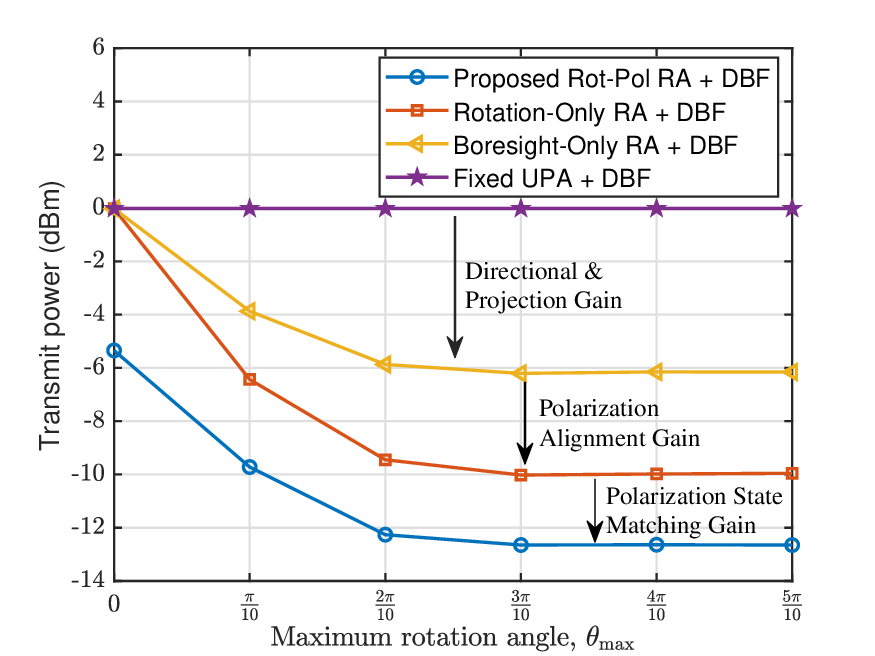}
  \caption{Transmit power versus maximum rotation angle, $\theta_{\max}$}
  \label{figure7}
  \vspace{-0.5em}
\end{figure}

Fig.~\ref{figure7} illustrates the impact of the maximum rotation angle $\theta_{\max}$ on the transmit power, with the directivity factor fixed at $p=2$. First, it is observed that the transmit power of the Fixed UPA + DBF schemes remains invariant with $\theta_{\max}$, as it does not possess rotation capabilities. In contrast, for the Proposed Rot-Pol RA + DBF and Rotation-Only RA + DBF schemes, the transmit power decreases monotonically as $\theta_{\max}$ increases and eventually plateaus. This saturation occurs when $\theta_{\max}$ exceeds approximately $70^\circ$, which corresponds to the maximum user angle set in the simulation, indicating that the antennas can fully cover the user distribution range. Notably, the relative performance of these schemes is consistent with the gain mechanisms revealed by our analysis in Section \uppercase\expandafter{\romannumeral3}. The initial power reduction from the Fixed UPA to the Boresight-Only scheme corresponds to the directional and projection Gain, unlocked by spatially steering the antenna boresight to maximize the effective aperture towards users. The further gain from Boresight-Only to Rotation-Only isolates the polarization direction alignment gain, which enables a better polarization orientation match between the transmitted field and the receive combiner. Finally, the performance leap from the Rotation-Only to the Proposed scheme represents the polarization state matching gain. This gain arises from utilizing polarization reconfiguration to compensate for channel depolarization, addressing the limitation that rotation alone can only adjust the orientation but cannot alter the polarization state. Additionally, at the specific point of $\theta_{\max}=0$, the performance of both the Boresight-Only and Rotation-Only schemes degrades to that of the Fixed UPA scheme, as the rotational DoFs are fully locked. In contrast, the Proposed scheme maintains a significant power advantage, validating the effectiveness of polarization reconfiguration even in the absence of 3D rotation.

\begin{figure}[t!]
  \centering
  \includegraphics[width=0.48\textwidth]{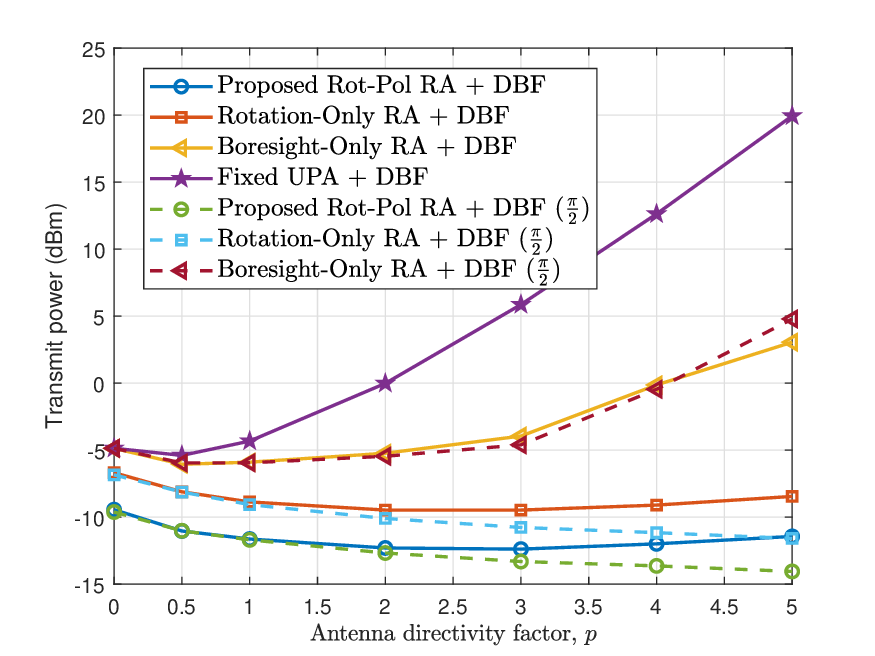}
  \caption{Transmit power versus antenna directivity factor, $p$}
  \label{figure4}
  \vspace{-0.5em}
\end{figure}

Fig. \ref{figure4} investigates the impact of the antenna directivity factor $p$ on the total transmit power. It is observed that when the maximum rotation angle is limited (e.g., $\theta_{\max} = \pi/5$), the transmit power for all schemes demonstrates an initial decline followed by a subsequent rise. Specifically, for the Fixed UPA + DBF scheme, increasing $p$ initially reduces the power by enhancing the peak directional gain $G_0$. However, as $p$ continues to rise, the power increases significantly. This is because the Fixed UPA scheme cannot mechanically steer its boresight to track the user. Consequently, as the beam narrows, the gain loss caused by angular misalignment exacerbates across all propagation paths, rapidly outweighing the benefits of the increased peak gain. Similarly, for the Proposed and Rotation-Only schemes with a limited rotation angle (e.g., $\theta_{\max}=\pi/5$), a non-monotonic trend is observed. This is primarily attributed to the rotation constraint, which prevents the antennas from steering their boresight towards users located outside the rotation range.  For these out-of-range users, the loss in the system arises from two main reasons: the intended LoS component is attenuated by boresight misalignment, while the increasingly narrow beam acts as a spatial filter that also suppresses the auxiliary NLoS components that would otherwise contribute to the effective channel. To validate this analysis, we further plot the performance curves with an extended rotation range of $\theta_{\max}=\pi/2$ (shown as dashed lines). It is observed that with sufficient rotation capability, the transmit power decreases monotonically with $p$. 
This is because the extended rotation range enables accurate boresight alignment with the dominant LoS direction, thereby recovering the directional gain and compensating for the misalignment-induced loss observed under a limited rotation range. As a result, the proposed architecture can fully exploit the high peak gain of narrow-beam antennas when the rotation capability is sufficient. However, for the Boresight-Only scheme, although it significantly outperforms the Fixed UPA scheme thanks to boresight alignment, its performance is not monotonic in $p$ even when $\theta_{\max}=\pi/2$. The reason is that while this scheme exploits the increasing directional gain by steering the boresight, it cannot actively control the polarization orientation. Therefore, the additional directional gain brought by a larger $p$ is often insufficient to offset the significant power loss caused by the uncorrected polarization mismatch, leading to a non-monotonic trend. This underscores the importance of accounting for polarization matching in rotatable antenna systems, since boresight alignment alone does not fully capture the gains enabled by antenna rotation.

\begin{figure}[t!]
  \centering
  \includegraphics[width=0.48\textwidth]{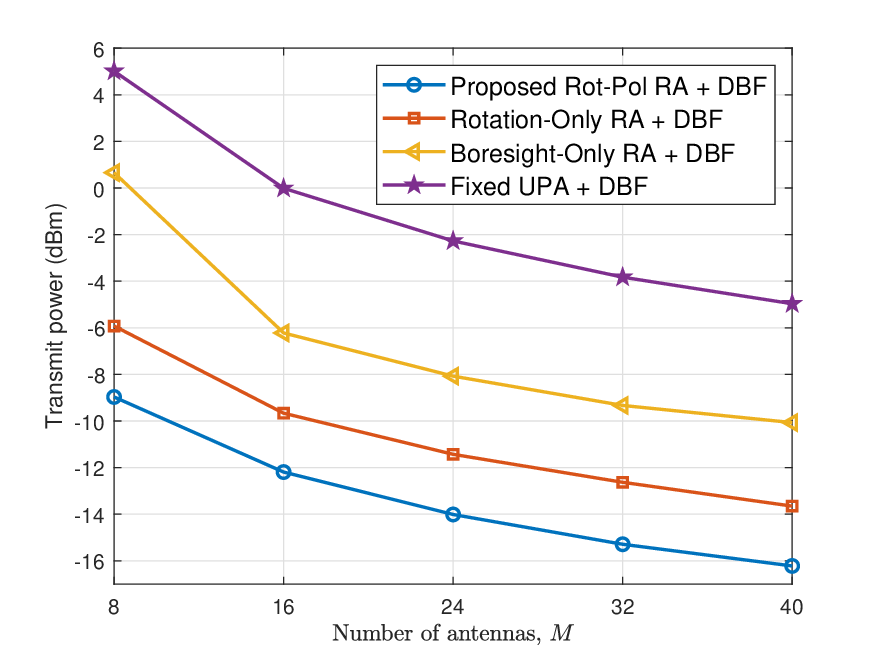}
  \caption{Transmit power versus number of antennas, $M$}
  \label{figure5}
  \vspace{-1em}
\end{figure}
\begin{figure}[t!]
  \centering
  \includegraphics[width=0.48\textwidth]{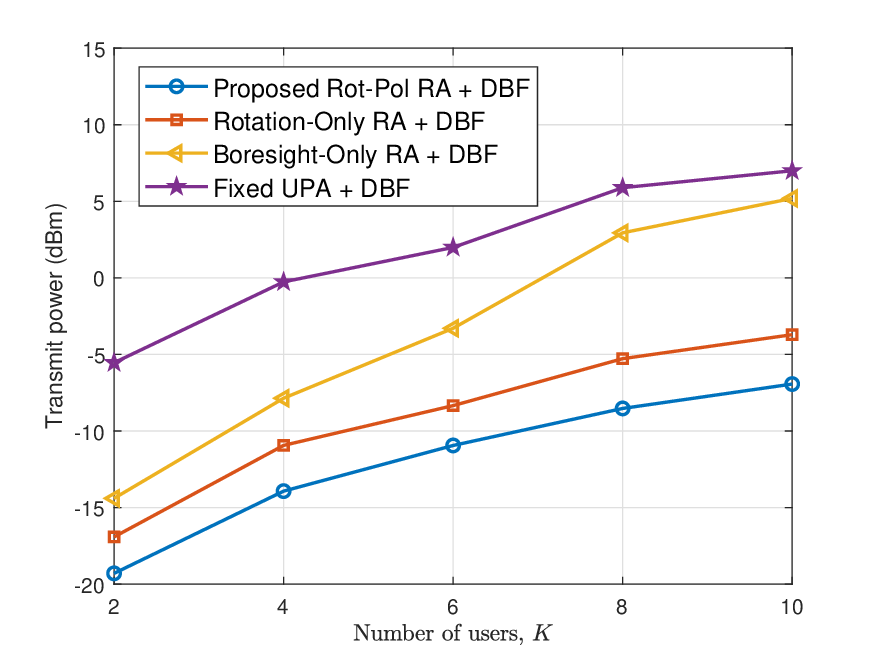}
  \caption{Transmit power versus number of users, $K$}
  \label{figure6}
  \vspace{-1em}
\end{figure}
Fig.~\ref{figure5} investigates the impact of the number of BS antennas $M$ on the total transmit power, with $p=2$ and $\bar{R}=2$ bps/Hz. We consider various UPA configurations, ranging from $4\times 2$ ($M=8$) to $4\times 10$ ($M=40$). It is observed that the transmit power for all schemes decreases monotonically as the number of antennas increases. This is because a larger antenna array offers higher array gains and improved spatial resolution, enabling more energy-efficient beamforming. Notably, the Proposed Rot-Pol RA + DBF scheme consistently achieves the lowest power consumption across all antenna configurations, maintaining a significant performance gap over the benchmarks.

Furthermore, Fig.~\ref{figure6} illustrates the transmit power versus the number of users $K$. As expected, the transmit power rises with the number of users. This increase is attributed to two factors: first, the BS must allocate additional power to satisfy the rate requirements of more users; second, the inter-user interference becomes more severe as $K$ increases, consuming extra DoFs for interference suppression. Despite the increased difficulty, the Proposed Rot-Pol RA + DBF scheme continues to outperform the benchmarks. Collectively, Fig.~\ref{figure5} and Fig.~\ref{figure6} validate the scalability and robustness of the proposed joint design, demonstrating its effectiveness in both large-scale MIMO setups and dense user scenarios.


\section{Conclusion}
In this paper, we have proposed a novel RA-enabled MU-MISO system where each antenna supports simultaneous 3D rotation and polarization reconfiguration. Through theoretical analysis in LoS scenarios, we have provided insights into the sources of the performance gains offered by the proposed RA architecture. Furthermore, for general multi-user multipath scenarios, we formulated a joint transmit power minimization problem and have solved it by developing an AO algorithm incorporating SDP and RCG techniques. Finally, simulation results have demonstrated that the proposed RA architecture significantly outperforms rotation-only and boresight-only benchmarks, validating the power savings achieved by exploiting the DoFs in both spatial and polarization domains.

{\appendix}
\section*{Appendix A: Euclidean Gradient for $\mathbf{R}_m$}
To implement the RCG algorithm, we derive the Euclidean gradient of the objective function $\mathcal{J}_R$ with respect to the rotation matrix $\mathbf{R}_m$. By applying the chain rule, the gradient is expressed as:
\begin{equation}
    \nabla_{\mathbf{R}_m} \mathcal{J}_R = \sum_{k=1}^{K} \!\omega_k \!\sum_{l=0}^{L} 2 \mathfrak{R} \left\{ \frac{\partial \gamma_k}{\partial h_{m,k,l}} \!\frac{\partial h_{m,k,l}}{\partial \mathbf{R}_m} \right\} \!+\! \lambda_1\! \nabla_{\mathbf{R}_m} P_{\text{angle}},\tag{43} \label{43}
\end{equation}
where $\omega_k= \frac{\partial (f_{\text{obj}} + \lambda_2 P_{\text{thresh}})}{\partial \gamma_k}$ represents the gradient of the objective function to the SINR
\begin{equation}
    \omega_k = -\frac{e^{-\mu \gamma_k}}{\sum_{j=1}^{K} e^{-\mu \gamma_j}} - 2\lambda_2 \cdot \text{sp}_{\alpha}(\iota_k - \gamma_k) \cdot \frac{e^{\alpha(\iota_k - \gamma_k)}}{1+e^{\alpha(\iota_k - \gamma_k)}}.\tag{44} \label{44}
\end{equation}
Moreover, we define $\Xi_{m,k}=\frac{\partial \gamma_k}{\partial h_{m,k,l}}$ as the partial derivative of the SINR with respect to the channel scalar $h_{m,k}$:
\begin{equation}
    \Xi_{m,k} = \frac{\partial \gamma_k}{\partial h_{m,k}} = \frac{I_k (\mathbf{h}_k^H \mathbf{w}_k) w_{k,m}^* - S_k \sum_{j \neq k} (\mathbf{h}_k^H \mathbf{w}_j) w_{j,m}^*}{I_k^2},\tag{45} \label{45}
\end{equation}
where $S_k = |\mathbf{w}_k^H \mathbf{h}_k|^2$ is the signal power, $I_k$ is the total interference-plus-noise power, and $w_{k,m}$ is the $m$-th element of $\mathbf{w}_k$. As for the term $\frac{\partial h_{m,k,l}}{\partial \mathbf{R}_m}$, let $\alpha_{m,k,l} = \sqrt{\beta_{m,k,l}} e^{-j \frac{2\pi}{\lambda} d_{m,k,l}}$ denote the complex path coefficient, which incorporates both the large-scale fading and the directional gain $(\mathbf{f}_{m,l}^T \mathbf{R}_m \mathbf{e}_z)^p$. Using the product rule, we derive:

\begin{equation}
    \frac{\partial h_{m,k,l}}{\partial \mathbf{R}_m} = \alpha_{m,k,l} \left( \mathbf{\Phi}_{m,k,l}^{\text{pol}} + \mathbf{\Phi}_{m,k,l}^{\text{gain}} \right).\tag{46} \label{46}
\end{equation}
Here, $\mathbf{\Phi}_{m,k,l}^{\text{pol}}$ and $\mathbf{\Phi}_{m,k,l}^{\text{gain}}$ are respectively given by 
\begin{align}
    \mathbf{\Phi}_{m,k,l}^{\text{pol}} &= \mathbf{Z}_{m,l} \mathbf{M}_l^T \mathbf{Q}_{k,l}^T \mathbf{u}_k^* (\mathbf{E} \mathbf{v}_m)^T, \tag{47} \label{47} \\
    \mathbf{\Phi}_{m,k,l}^{\text{gain}} &= p (\mathbf{f}_{m,l}^T R_m \mathbf{e}_z)^{-1} (\mathbf{u}_k^H \mathbf{Q}_{k,l} \mathbf{M}_l \mathbf{P}_{m,l} \mathbf{v}_m) \mathbf{f}_{m,l} \mathbf{e}_z^T.\tag{48} \label{48}
\end{align}
Finally, the gradient of the angle penalty term is $\nabla_{\mathbf{R}_m} P_{\text{angle}} = -\kappa_m \mathbf{e}_z \mathbf{e}_z^T$, where the scalar coefficient $\kappa_m$ is $\kappa_m = 2 \text{sp}_{\alpha}(\Delta_m) \cdot \frac{e^{\alpha \Delta_m}}{1+e^{\alpha \Delta_m}}$,
with $\Delta_m = \cos\theta_{\max} - \mathbf{e}_z^T \mathbf{R}_m \mathbf{e}_z$ representing the violation of the elevation angle constraint.

\section*{Appendix B: Euclidean Gradient for $\mathbf{v}_m$}
The Euclidean gradient of the objective function $\mathcal{J}_V$ with respect to the transmit polarization vector $\mathbf{v}_m$ is derived as the partial derivative with respect to $\mathbf{v}_m^*$. By applying the chain rule, the gradient is expressed as
\begin{equation}
    \nabla_{\mathbf{v}_m} \mathcal{J}_V = \sum_{k=1}^{K} \omega_k \sum_{l=0}^{L} \Xi_{m,k}^* \cdot \mathbf{\Psi}_{m,k,l},\tag{49} \label{49}
\end{equation}
where $\omega_k$ is defined in Appendix A. Note that $\Xi_{m,k}^*$ is the conjugate of the term $\Xi_{m,k}$ defined in Appendix A, corresponding to the derivative $\frac{\partial \gamma_k}{\partial h_{m,k}^*}$.
The vector term $\mathbf{\Psi}_{m,k,l}$ represents the gradient of the conjugate channel component with respect to $\mathbf{v}_m^*$, given by
\begin{equation}
    \mathbf{\Psi}_{m,k,l} =\frac{\partial h_{m,k,l}^*}{\partial \mathbf{v}_m^*} = \alpha_{m,k,l}^* \mathbf{P}_{m,k,l}^T \mathbf{M}_{m,k,l}^H \mathbf{Q}_{m,k,l}^T \mathbf{u}_k,\tag{50} \label{50}
\end{equation}
where $\alpha_{m,k,l}$ is defined in Appendix A.
\section*{Appendix C: Euclidean Gradient for $\mathbf{u}_k$}

The Euclidean gradient of $\mathcal{J}_U$ with respect to the receive polarization vector $\mathbf{u}_k$ is derived as the partial derivative with respect to $\mathbf{u}_k^*$. Since $\mathbf{u}_k$ is specific to the $k$-th user, the gradient does not involve a summation over users:
\begin{equation}
    \nabla_{\mathbf{u}_k} \mathcal{J}_U = \omega_k \sum_{m=1}^{M} \Xi_{m,k} \sum_{l=0}^{L} \mathbf{\Upsilon}_{m,k,l},\tag{51} \label{51}
\end{equation}
where $\omega_k$ and $\Xi_{m,k}$ are defined in Appendix A. 
The vector term $\mathbf{\Upsilon}_{m,k,l}$ represents the gradient of the channel component $h_{m,k,l}$ with respect to $\mathbf{u}_k^*$, given by:
\begin{equation}
    \mathbf{\Upsilon}_{m,k,l}=\frac{\partial h_{m,k,l}}{\partial \mathbf{u}_k^*} = \alpha_{m,k,l} \mathbf{Q}_{m,k,l} \mathbf{M}_{m,k,l} \mathbf{P}_{m,k,l} \mathbf{v}_m.\tag{52} \label{52}
\end{equation}
\bibliographystyle{IEEEtran}
\bibliography{main.bib}
\end{document}